\documentclass[a4paper, 12pt, fleqn]{article}

\usepackage{authblk}
\usepackage[a4paper,top=2cm,bottom=2cm,left=2cm,right=2cm,marginparwidth=1.75cm]{geometry}
\usepackage[colorlinks=true, allcolors=blue]{hyperref}

\usepackage{natbib}
\usepackage[english]{babel}
\usepackage[utf8x]{inputenc}
\usepackage[T1]{fontenc}
\usepackage{lmodern}

\usepackage[title]{appendix}
\usepackage{xcolor}
\usepackage{textcomp}
\usepackage{manyfoot}
\usepackage{enumitem}
\usepackage{listings}
\usepackage{placeins}

\usepackage{amsmath, amssymb, amsfonts, amsthm, mathtools, nicefrac, amscd, latexsym, mathrsfs}
\mathtoolsset{showonlyrefs}
\usepackage{MnSymbol}

\usepackage{graphicx}
\usepackage[format=hang, font=small, labelfont=bf]{caption}
\usepackage{subcaption}

\usepackage{tabularx, fix-cm, array}
\usepackage{multirow, multicol}
\usepackage{booktabs}

\usepackage[algo2e, ruled]{algorithm2e}%

\newtheorem{theorem}{Theorem}
\newtheorem{proposition}[theorem]{Proposition}

\providecommand{\keywords}[1]{\small\textbf{{Keywords.}} #1}

\newcommand{\dirsim}{Figures}
\newcommand{\direx}{\dirsim}
\newcommand{\botrule}{\bottomrule}

\newcommand*{\bZ}{{\mathbf{Z}}}
\newcommand*{\bz}{{\mathbf{z}}}

\newcommand*{\bV}{{\mathbf{V}}}
\newcommand*{\bv}{{\mathbf{v}}}

\newcommand*{\bY}{{\mathbf{Y}}}

\newcommand*{\bx}{{\mathbf{x}}}
\newcommand*{\bo}{{\mathbf{o}}}
\newcommand*{\bX}{{\mathbf{X}}}

\newcommand*{\btheta}{{\boldsymbol{\theta}}}
\newcommand*{\bpsi}{{\boldsymbol{\psi}}}
\newcommand*{\tbeta}{{\widetilde{\beta}}}
\newcommand*{\bbeta}{{\boldsymbol{\beta}}}
\newcommand{\bBeta}{{\boldsymbol{B}}}
\newcommand{\bSigma}{{\boldsymbol{\Sigma}}}

\newcommand*{\bm}{{\mathbf{m}}}
\newcommand*{\bS}{{\mathbf{S}}}
\newcommand{\bI}{\mathbf{I}}
\newcommand{\bG}{\mathbf{G}}
\newcommand{\bH}{\mathbf{H}}
\newcommand{\bJ}{\mathbf{J}}
\newcommand{\bA}{\mathbf{A}}

\renewcommand*{\d}{\mathrm{d}}
\newcommand*{\esp}[2][]{%
	\mathbb{E}_{#1}\mathopen{}\mathclose\bgroup\left[#2%
	\aftergroup\egroup\right]%
}
\newcommand*{\is}[2][]{%
	\widehat{\mathbb{E}}_{#1}\mathopen{}\mathclose\bgroup\left[#2%
	\aftergroup\egroup\right]%
}
\newcommand*{\var}[2][]{%
	\mathbb{V}{\rm ar}_{#1}\mathopen{}\mathclose\bgroup\left[#2%
	\aftergroup\egroup\right]%
}

\DeclareMathOperator*{\argmax}{arg\,max\,}
\newcommand*{\tr}{\mathrm{tr}}

\renewcommand{\mid}{\;\ifnum\currentgrouptype=16 \middle\fi|\;}

\newcommand{\cl}{c\ell}

\newcommand{\Var}{\mathbb{V}{\rm ar}}

\begin{document}

\title{Composite likelihood inference for the Poisson log-normal model}
\date{}
\author[1, 2]{Julien Stoehr}
\author[3]{St\'ephane Robin}
\affil[1]{CEREMADE, Universit\'e Paris-Dauphine, Universit\'e PSL, CNRS, 75016 Paris, France.}
\affil[2]{Universit\'e Paris-Saclay, INRAE, AgroParisTech, UMR MIA Paris-Saclay, 91120 Palaiseau, France.}
\affil[3]{Sorbonne Universit\'e, CNRS, Laboratoire de Probabilit\'es, Statistique et Mod\'elisation, 75005 Paris, France.}

\maketitle

\begin{abstract}
The Poisson log-normal model is a latent variable model that provides a generic framework for the analysis of multivariate count data. Inferring its parameters can be a daunting task since the conditional distribution of the latent variables given the observed ones is intractable. For this model, variational approaches are the golden standard solution as they prove to be computationally efficient but lack theoretical guarantees on the estimates. Sampling-based solutions are quite the opposite. 
We first define a Monte Carlo EM algorithm that can achieve maximum likelihood estimators, but that is computationally efficient only for low-dimensional latent spaces.
We then propose a novel inference procedure combining the EM framework with composite likelihood and importance sampling estimates. 
The algorithm preserves the desirable asymptotic properties of maximum likelihood estimators while circumventing the high-dimensional integration bottleneck, thus maintaining computational feasibility for moderately large datasets.
This approach enables grounded parameter estimation, confidence intervals, and hypothesis testing. Application to the Barents Sea fish dataset demonstrates the algorithm capacity to identify significant environmental effects and residual interspecies correlations.

\vspace{\baselineskip}
\noindent\keywords{Composite likelihood, importance sampling, Monte Carlo EM algorithm, multivariate count data}
\end{abstract}

\section{Introduction}

\subsection{The Poisson log-normal model} \label{subsec:PLN}

The multivariate Poisson log-normal model \citep[PLN:][]{AiH89} is a generic model for the joint distribution of count data that provides a flexible framework to account for dependencies between the different counts. It also exhibits overdispersion, a common feature in several application domains such as ecology, epidemiology and genomics, to name a few. 
It is a special instance of a latent variable model accounting for covariates. 
From a general point of view, a latent variable model aims to describe the variation of a response variable $\bY$ conditionally to unobserved (\textit{i.e.}, {\sl latent}) variables $\bZ$ and, possibly, observed covariates $\bx$. 
As a typical example in synecology (or community ecology), one may consider $n$ sites (indexed by $1 \leq i \leq n$) and $p$ animal species (indexed by $1 \leq j \leq p$) and denote by $Y_{ij}$ the number of individuals from species $j$ observed in site $i$. 
Since the number of species observed in different areas may vary due to environmental factors, it is further assumed that a $d$-dimensional vector of covariates (descriptors) $\bx_i$ describing the environmental conditions is recorded for each site $i$.
In order to model the dependencies between the species, the PLN model assumes that, for each site $i$, the $p$-dimensional observations relate to a latent $p$-dimensional Gaussian random vector $\bZ_i = (Z_{ij})_{1 \leq j \leq p}$ with covariance matrix $\bSigma$ (typically non-diagonal).

Namely, the observed counts $Y_{ij}$ conditionally on $\bZ_i$ are independent and distributed according to a Poisson distribution with rate $\exp(o_{ij} + \bx_i^\top\bbeta_j + Z_{ij})$: 
\begin{equation}
\label{eqn:pln}
    \begin{aligned}
	\{\bZ_i\}_{1 \leq i \leq n} \textit{ i.i.d.}:  & & \bZ_i & \sim \mathcal{N}(\mathbf{0}, \bSigma),
	\\
  	\{Y_{ij}\}_{1 \leq i \leq n, 1 \leq j \leq p}  \text{ independent} \mid \{\bZ_i\}_{1 \leq i \leq n}:  & & 
		Y_{ij} \mid Z_{ij} & \sim \mathcal{P}\left(\exp(o_{ij} + \bx_i^\intercal \bbeta_j + Z_{ij})\right), 
    \end{aligned}
\end{equation}
where $o_{ij}$ is an offset parameter, $\bbeta_j = [\beta_{1j} \dots \beta_{dj}]^\top \in \mathbb{R}^d$ is a species-specific vector of regression parameters
and where the covariance matrix $\boldsymbol{\Sigma} = [\sigma_{jk}]_{1 \leq j, k \leq p}$ encodes the dispersion and dependency structure between the counts from a same site.

In this model, the regression coefficient $\beta_{\ell j}$ ($1 \leq \ell \leq d$, $1 \leq j \leq p$) is interpreted as the effect of covariate $\ell$ of the mean abundance of species $j$, whereas $\sigma_{jk}$ ($1 \leq j, k \leq p$) is the latent covariance coefficient between the abundances of species $j$ and $k$. 
Gathering the vectors of regression coefficients into a $d \times p$ matrix $\bBeta = [\bbeta_1\,\cdots\,\bbeta_p]$, we denote by $\btheta = (\bBeta, \bSigma)$ the unknown parameter of the PLN model.

\paragraph*{Barents Sea dataset} The application of interest is the study of the fish species from the Barents Sea. The dataset consists of the abundances of $p=30$ fish species in $n=89$ stations (sites) from the Barents Sea, collected between April and May 1997. We refer the reader to \cite{FNA06} for a more detailed description. For each sample, four covariates: the latitude and longitude of the station, as well as the temperature of the water and the depth, were recorded. Abundances were all obtained using the same experimental protocol, so no offset term is required in the model. The data are available from the {\tt PLNmodels} R package \citep{CMR21}. 

The main objective for the practitioner is to address two key questions, namely ($i$) which environmental covariates have a significant effect on each species under study? ($ii$) which pairs of species display a correlation that does not result from environmental variations? 
We thereby need an inference framework that provides statistically grounded estimators, specifically, consistent and asymptotically normal estimators, that enable quantifiable measures of uncertainty and rigorous hypothesis testing.

\subsection{Maximum likelihood and the expectation-maximization algorithm}

Maximum likelihood inference for latent variable models is usually not straightforward because the likelihood of the data $p_{\btheta}(\bY)$ results from a high-dimensional integration over the unobserved $\bZ$: $p_{\btheta}(\bY) = \int p_{\btheta}(\bY \mid\bZ) p_{\btheta}(\bZ) \d \bZ$. This problem can be circumvented using the celebrated expectation-maximization (EM) algorithm \citep{DLR77}, which relies on the evaluation of certain moments of the conditional distribution $p_{\btheta}(\bZ \mid \bY)$.
Importantly, the resulting estimate inherits the general properties of maximum likelihood estimates; in particular, its asymptotic variance can be evaluated using side products of the EM algorithm \citep{Lou82}.

Unfortunately, the conditional distribution $p_{\btheta}(\bZ \mid \bY)$ of the PLN model is intractable, as is often the case with even moderately complex latent models.
Consequently, estimating the parameters of such a model requires to resort to alternative strategies that can deal with this complex conditional distribution.
Such strategies can be cast into two main approaches: either approximate it or sample from it. The former approach can typically rely on variational approximations, whereas the latter gave rise to a series of stochastic versions of EM.

\paragraph*{Variational approximation of the EM algorithm}
The most used approach for the PLN model is arguably the variational EM (VEM) algorithm, which was developed by \cite{CMR18, CMR19}, and implemented in the R package {\tt PLNmodels}. 
Variational approximations \citep[see \textit{e.g.},][]{JaJ00, WaJ08, BKM17} have received significant attention in the last decade due to their computational efficiency and their (reasonable) ease of implementation. 
Nevertheless, although VEM algorithms usually demonstrate good empirical performance, the resulting estimate is not guaranteed, in general --- and specifically in the PLN case ---, to have any desirable statistical properties, such as consistency or asymptotic normality, and no measure of uncertainty or significance test can be provided to practitioners in the end.


\paragraph*{Monte-Carlo approximations of the EM algorithm}
Since the variational approach does not provide a suitable interpretation for the practitioner, it seems desirable to design a solution inheriting maximum likelihood properties.
An alternative way to address the intractability of the conditional distribution $p_{\btheta}(\bZ \mid \bY)$ is to resort to Monte Carlo EM (MCEM) algorithms \citep{wei1990}
specifically designed to deal with the E-step which cannot be performed analytically or is computationally cumbersome.
A Monte Carlo sampling method is then performed in the E-step,
such as important sampling \citep{BH1999, LC2001} or Markov Chain Monte Carlo \citep{Culloch1997, FM2003}, to provide estimates of the needed conditional moments. 
However, all these methods require to sample from the conditional distribution $p_{\btheta}(\bZ \mid \bY)$, which is time-consuming and can be poorly efficient, especially when the dimension of $\bZ$ is large, that is, when the number of species exceeds ten or so.

\paragraph*{Composite likelihood} 
The composite likelihood (CL) approach offers a way to overcome the dimensional curse when dealing with a moderately large number of species. It refers
to a broad variety of approaches where the likelihood of the data is replaced with a modified version of it. Such approaches trace back to \cite{Bes74} and \cite{Lin88}, who proposed an (almost) general definition, which consists in splitting the data into a series of (possibly overlapping) blocks and to replace the log-likelihood with a {convex combination} of log-likelihoods corresponding to each block of data.
Splitting the dataset into blocks allows dealing with large dimension data and/or complex dependency structure, while keeping desirable statistical properties \citep[see, e.g.][]{VRF11}. We will see that, in the presence of latent variables, the classical EM algorithm can be extended to deal with the composite likelihood, as well as its variational or Monte Carlo counterparts. 

\subsection{Contribution}

Our contribution is three-fold. 
Importantly, the proposed algorithm gives access to the asymptotic variance of the parameters and, consequently, to confidence intervals and tests of the parameters, as opposed to the available VEM algorithm.
Our second contribution is the use of composite likelihood inference, using blocks of species.
Indeed, beyond the intractability concern, 
we also need to develop a method that scales up to a few tens of species, likewise in the Barents Sea study where $p=30$.
This approach allows to consider latent variables living in a space with a smaller dimension (actually, the number of species in each block), so to make importance sampling efficient.
To the best of our knowledge, this results in the first EM algorithm 
combining importance sampling and composite likelihood to address inference on such a model.
Third, we show how to define a mixture proposal distribution that ensures finite variance on bounded test functions and controls the efficiency of the importance sampling steps.
The overall scheme thus allows to answer the key questions on the Barents Sea dataset, and, more generally, to analyse datasets from synecology or biogeography involving a moderately large number of species.

\paragraph*{Outline}
In Section \ref{sec:reminder}, we 
recall notions
and properties about both EM and VEM, essential for understanding the inference on the PLN model and its challenges.
We then describe, in Section \ref{sec:CL}, how to extend the EM algorithm to composite likelihood inference.
Specifically, we demonstrate the validity of the scheme.
In Section \ref{sec:ISEM}, we introduce the MCEM algorithm, based on importance sampling, which applies to both likelihood and composite likelihood inference.
The method requires selecting a collection of blocks of species that is discussed at the beginning of Section \ref{sec:illus}. Throughout this section, we also illustrate our algorithm on synthetic data. 
After numerically demonstrating the benefits of our approach over the existing variational solutions, we provide a thorough analysis and a conclusion regarding the Barents Sea dataset.

\section{Reminder on the inference for the Poisson log-normal model \label{sec:reminder}}

\subsection{Maximum likelihood inference using expectation-maximization} \label{sec:EM}

The inference of any incomplete data model can be tackled using the expectation-maximization (EM) algorithm \citep{DLR77}. The latter is an iterative procedure (Algorithm \ref{algo:EM}) based on maximizing the conditional expectation of the complete log-likelihood under the current estimates $\btheta^{(h)}$ of model parameters, defined for $\btheta \in \Theta$ by 
\begin{equation} \label{eqn:Q}
	Q(\btheta \mid \btheta^{(h)}) = 
	\esp[\btheta^{(h)}]{\log p_{\btheta}(\bY, \bZ) \mid \bY}.
\end{equation}
The algorithm generates a sequence that converges under regularity conditions to the maximum likelihood estimator \citep{wu1983, boyles1983}.

\begin{algorithm2e}[!h]
\caption{EM} \label{algo:EM}
  \Repeat{convergence}{
  {\bf E-step:} compute the moments of the conditional distribution $p_{\btheta^{(h)}}(\bZ \mid \bY)$ required to evaluate $Q(\btheta \mid \btheta^{(h)})$\; 
  {\bf M-step:} update the parameter estimate as:
    \begin{equation*}
      \btheta^{(h+1)} = \arg\max_{\btheta} Q(\btheta \mid \btheta^{(h)})\,;
    \end{equation*}
  }
\end{algorithm2e}

\paragraph*{Asymptotic variance of the maximum likelihood estimates}
Because EM is a maximum likelihood method, under fairly general regularity conditions, the resulting estimators are asymptotically unbiased, Gaussian, with known asymptotic variance.
The observed Fisher information matrix based on a $n$-sample of observations $(\bY_1, \ldots, \bY_n)$ writes as
\begin{equation}
\label{eqn:obs-fim}
\widehat{\bI}_n(\btheta) = \frac{1}{n}\sum_{i = 1}^{n}{\nabla_{\btheta} \log p_{\btheta}(\bY_i) \{\nabla_{\btheta} \log p_{\btheta}(\bY_i)\}^\top}. 
\end{equation}
Since integrating over the latent space might be cumbersome, \cite{Lou82} provides a reformulation of the score function for incomplete data models, which makes the exact computation of $\widehat{\bI}_n(\btheta)$ possible on some models using side information resulting from EM inference. More specifically, 
 \begin{equation} 
 \label{eqn:louis-1}
 \nabla_{\btheta} \log p_{\btheta}(\bY) = \esp[\btheta]{\nabla_{\btheta}\log p_{\btheta}(\bY, \bZ) \mid \bY}.
 \end{equation}
We refer the reader to Appendix \ref{app:fim} for an alternative estimator of the Fisher information matrix.

\paragraph*{Objective function for the PLN model}
The complete likelihood associated with the PLN model writes (up to constants with respect to the parameters)
\begin{align*}
  \log p_{\btheta}(\bY, \bZ)
  & = \log p_{\bSigma}(\bZ) + \log p_{\bBeta}(\bY \mid \bZ)  \\
  & = - \frac{n}{2} \log \vert \bSigma\vert - \frac{1}{2} \sum_{i=1}^n \bZ_i^\top \bSigma^{-1} \bZ_i  + \sum_{i=1}^n \sum_{j=1}^p \left\lbrace Y_{ij}(o_{ij} + \bx_i^\top\bbeta_j + Z_{ij})  -\exp(o_{ij} + \bx_i^\top\bbeta_j + Z_{ij})  \right\rbrace.
\end{align*}
Therefore, to perform the E-step of Algorithm \ref{algo:EM} and to compute \eqref{eqn:louis-1},  we need to evaluate , for all $1 \leq i \leq n$ and $1 \leq j \leq p$, the conditional moments:
\begin{equation} 
\label{eqn:condMoments}
	\esp[\btheta^{(h)}]{Z_{ij} \mid Y_{ij}},
	\qquad 
	\esp[\btheta^{(h)}]{\exp(Z_{ij}) \mid Y_{ij}} 
	\qquad \text{and} \qquad
	\esp[\btheta^{(h)}]{\bZ_i^\top \bSigma^{-1} \bZ_i \mid \bY_i}.
\end{equation}
Unfortunately, one major limitation of the PLN model is that the conditional distribution $p_{\btheta}(\bZ \mid \bY)$ is intractable, and so are these conditional moments, even in the one-dimensional case.

\subsection{Variational expectation-maximization \label{sec:VEM}}

Variational inference circumvents the lack of a closed-form by replacing $p_{\btheta}(\bZ \mid \bY)$ with a surrogate distribution $q_\bpsi(\bZ)$, chosen within a certain class of distributions, parametrized by $\bpsi$. 
The variational EM (VEM) algorithm then amounts at maximizing a lower bound of the log-likelihood,
namely $\log p_\btheta(\bY) - \mathrm{KL}\left[q_\bpsi(\bZ) \Vert p_\btheta(\bZ \mid \bY)\right]$, where KL stands for the Kullback-Leibler divergence between $q_\bpsi$ and $p_\btheta(\bZ \mid \bY)$.
The VEM alternates the update of the {\sl variational} parameter $\bpsi$, with that of the model parameter $\btheta$. 
The reader may refer to \cite{Jaa01} or \cite{WaJ08} for a general introduction.

\paragraph*{Variational family for the PLN model}
For the PLN model, \cite{CMR18, CMR19} proposed using the class of Gaussian distributions. Because the sites $1 \leq i \leq n$ are supposed to be independent, each conditional distribution $p_\btheta(\bZ_i \mid \bY_i)$ is approximated with a normal distribution $\mathcal{N}(\bm_i, \bS_i)$ with diagonal covariance matrix $\bS_i$. The variational parameter is hence $\bpsi = \{(\bm_i, \bS_i)\}_{1 \leq i  \leq n}$.

\paragraph*{Variance of the variational estimates for the PLN model}
An important feature of variational inference is that the resulting estimator of $\btheta$ does not enjoy the same properties as the regular maximum-likelihood estimator; in particular, there are no general results ensuring consistency or asymptotic normality. In summary, variational methods computational efficiency comes at the price of a loss of statistical guarantees.
The {\tt PNmodels} R package incorporates two solutions to approximate the unknown (asymptotic) variance of the variational estimator. 
The first (referred to as \textit{variational} in the package) 
uses the evidence lower bound (ELBO) in place of the log-likelihood and estimates the variance using the observed Fisher information matrix \eqref{eqn:obs-fim} where the score function is derived from the ELBO instead of the log-likelihood.
Alternatively, the package provides the jackknife (JK) estimator \citep{EfS81}, which requires running the VEM algorithm $n$ times, leaving one observed site out in each run.
Due to this leave-one-out strategy, note that each of these $n$ runs converges faster than the scheme on the entire dataset.


\section{Composite likelihood inference \label{sec:CL}}

As an alternative to variational methods, we consider in Section \ref{sec: ML-ISEM} Monte Carlo estimation of the conditional moments \eqref{eqn:condMoments}. 
To overcome sampling issues in moderately high-dimensional latent spaces, we turn to composite likelihood (CL), which we briefly present below.
It is another admissible contrast for grounded statistical inference where the observed variables $\bY$ are distributed into $C$ (possibly overlapping) blocks $\{\mathcal{C}_b\}_{1 \leq b \leq C}$ of data, and is defined as
\begin{equation} \label{eqn:cl}
  \cl_{\btheta}(\bY) = \sum_{b=1}^C \lambda_b \; \log p_{\btheta}(\bY^{(b)}), 
\end{equation}
where $\lambda_b$ is the weight associated with block $\mathcal{C}_b$ and $\bY^{(b)}$ is the dataset reduced to block $\mathcal{C}_b$. Most of the time, all blocks are chosen to have the same size $k$, pairwise composite likelihood referring to $k = 2$. 

The maximum composite likelihood estimate is then defined as $\widehat{\btheta}_{\cl} = \argmax_{\btheta} \cl_{\btheta}(\bY)$. Importantly, $\widehat{\btheta}_{\cl}$ enjoys properties similar to those of the maximum likelihood estimator, namely consistency and asymptotic normality \citep[see][for a general review]{VRF11}, although, possibly, with higher asymptotic variance \citep{ZhJ05, XRX16}. These properties actually derive from the more general properties of M-estimators \citep[][Chap. 5]{vdV98}. 

\paragraph*{Composite likelihood inference using EM \label{sec:CL-EM}}
Composite likelihood can be used for the inference of an incomplete data model, using an adaptation of the EM Algorithm \ref{algo:EM} \citep[see, e.g.][]{VHS05, BCK15}. 
Following \eqref{eqn:cl}, the EM objective is, for each block, the expectation of $\log p_{\btheta}(\bY^{(b)}, \bZ)$ with respect to the conditional distribution of the whole latent variables $p_\btheta(\bZ\mid\bY^{(b)})$.
However, models such as the PLN model allow to distribute the latent variables $\bZ$ into the same $C$ blocks, so that we only need to deal with the conditional distribution of a subset of random variables, namely: $p_{\btheta}(\bZ \mid \bY^{(b)}) = p_\btheta(\bZ^{(b)}\mid\bY^{(b)})$.
One may then design the CL-EM Algorithm \ref{algo:CL-EM}.

\begin{algorithm2e}[!h]
\caption{CL-EM}
\label{algo:CL-EM}
  \Repeat{convergence}{
  {\bf CL-E-step:} compute the moments of each of the $C$ conditional distribution $p_{\btheta^{(h)}}(\bZ^{(b)} \mid \bY^{(b)})$ required to evaluate
  \begin{equation*}
      Q_{\cl}(\btheta \mid \btheta^{(h)}) =  \sum_{b=1}^C \lambda_b \esp[\btheta^{(h)}]{\log p_{\btheta}(\bY^{(b)}, \bZ^{(b)}) \mid \bY^{(b)}};
    \end{equation*}
  {\bf CL-M-step:} update the parameter estimate as:
    \begin{equation*}
      \btheta^{(h+1)} = \arg\max_{\btheta} Q_{\cl}(\btheta \mid \btheta^{(h)})\,;
    \end{equation*}
  }
\end{algorithm2e}

Algorithm \ref{algo:CL-EM} enjoys the same general property as the EM Algorithm \ref{algo:EM}. The proof follows the same line as this for EM and is given in Appendix \ref{app:proof}. Note that a similar result was obtained by \cite{VHS05} in the specific case of pairwise composite likelihood.

\begin{proposition} \label{prop:CL-EM}
If $p_{\btheta}(\bZ \mid \bY^{(b)}) = p_\btheta(\bZ^{(b)}\mid\bY^{(b)})$, using Algorithm \ref{algo:CL-EM} yields a sequence $(\btheta^{(h)})_{h\in\mathbb{N}}$ such that $\cl_{\btheta^{(h+1)}}(\bY) \geq \cl_{\btheta^{(h)}}(\bY)$.
\end{proposition}

\paragraph*{Asymptotic variance of the maximum composite likelihood estimates}
The asymptotic variance of the maximum composite likelihood estimator $\widehat{\btheta}_{\cl}$
is given by the inverse of the so-called Godambe information matrix \citep{VRF11} defined for all $\btheta\in\Theta$ as
\begin{equation} \label{eqn:godambe}
\bG(\btheta) = \bH(\btheta)\bJ(\btheta)^{-1}\bH(\btheta)
\quad\text{where}\quad
\begin{cases}
\bJ(\btheta) & = \var[\btheta]{\nabla_{\btheta} \cl(\btheta)},
\\
\bH(\btheta) & = -\esp[\btheta]{\nabla^2_{\btheta}\cl(\btheta)}.
\end{cases}
\end{equation}
where the expectation and the variance are taken with respect to $\bY$.
In the general framework, the computation of $\bG(\btheta)$ can be achieved using solely the gradient of the log-marginal on each block.
\begin{proposition}
For all $\btheta \in \Theta$, the matrix $\bH$ writes as a convex sum of Fisher information matrices of each block
\begin{equation*}
\bH(\btheta) = \sum_{b = 1}^C \lambda_b \esp[\btheta]{\nabla_{\btheta} \log p_{\btheta}(\bY^{(b)}) \{\nabla_{\btheta} \log p_{\btheta}(\bY^{(b)})\}^\top}^{\blacksquare}.
\end{equation*}
where $\bA^{\blacksquare}$ stands for a $k\times k$ matrix $\bA$ that has been embedded into a $p\times p$ matrix $\mathbf{M} = (M_{j\ell})$ with $M_{j\ell}$ taking value 0 if the pair $(j,\ell)$ does not belong to the possible pair of species of block $\mathcal{C}_b$ and being equal to the coefficient of $\bA$ related to the pair $(j,\ell)$ otherwise.
\end{proposition}
This expression stems from the well known identity for regular models 
\begin{equation}
\label{eqn:id-fim}
\var[\btheta]{\nabla_{\btheta} \log p_{\btheta}(\bY)} = - \esp[\btheta]{\nabla^2_{\btheta} \log p_{\btheta}(\bY)}, 
\end{equation}
holds for each block of data $\bY^{(b)}$ and that the score of the likelihood on a given block has a zero expectation. 

Like in the regular likelihood framework, one can derive a sample variance estimator for $\bJ(\btheta)$ and a sample mean estimator for $\bH(\btheta)$. 
For both estimators, $\{\nabla_{\btheta} \log p_{\btheta}(\bY^{(b)})\}_{1 \leq b \leq C}$ terms can be computed applying Louis' formula \eqref{eqn:louis-1}. 
Louis' formula \eqref{eqn:louis-2} could also be used in the context of composite likelihood to derive control variate estimators.

\paragraph*{Model selection} \label{sec:CL-select}
Model selection procedure also exists in the context of composite likelihood inference, and analogues of the Akaike information criterion  \citep[AIC,][]{VaV05} and Bayesian information criterion \citep[BIC,][]{GaS10} criteria have been derived in this context. The latter reference proves the consistency of the composite likelihood BIC criterion, defined as
\begin{equation} \label{eq:cl-bic}
  BIC = \cl_{\btheta}(\bY) - \frac{\log(n)}2 \text{dim}(\btheta), 
  \quad 
  \text{where} \quad
  \text{dim}(\btheta) 
  = \tr[\bH(\btheta) \bG(\btheta)^{-1}].
\end{equation}
Observe that, in the regular likelihood context, both $\bH(\btheta) \bG(\btheta)^{-1} = \bJ(\btheta) \bH(\btheta)^{-1}$ reduce to the identity matrix and $\dim(\btheta)$ is the number of independent parameters, which yields the regular BIC formula.
In Section \ref{sec:CL-ISEM}, we will derive estimates of the matrices $\bH(\btheta)$, $\bJ(\btheta)$ and $\bG(\btheta)$, which will enable us to evaluate the composite likelihood BIC criterion.

\paragraph*{Composite formulation for the PLN model} 
In the context of the PLN model, the blocks will consist of subsets of $k$ species, that is $\mathcal{C}_b \subset \{1, \dots, p\}$ and $\bY^{(b)} = \{Y_{ij}\}_{1 \leq i \leq n, j \in \mathcal{C}_b}$.
We shall consider the same blocks of species for $\bZ$ as for $\bY$, that is $\bZ^{(b)} = \{Z_{ij}\}_{1 \leq i \leq n, j \in \mathcal{C}_b}$. 
Obviously, the conditional moments to be evaluated at the CL-E-step are the same than these given in \eqref{eqn:condMoments}, replacing $\bY$ (resp. $\bZ$) with $\bY^{(b)}$ (resp. $\bZ^{(b)}$) for each block $\mathcal{C}_b$, but with a further simplification on the parameter. The complete likelihood $p_\btheta(\bY^{(b)}, \bZ^{(b)})$ of block $\mathcal{C}_b$ does not depend on all the elements of $\btheta$. More specifically, it only involves $\btheta^{(b)} = (\bBeta^{(b)}, \bSigma^{(b)})$ where
\begin{equation*}
\bBeta^{(b)} = [\bbeta_j]_{j \in \mathcal{C}_b}, 
\quad\text{and}\quad
\bSigma^{(b)} = [\sigma_{jk}]_{(j, k) \in \mathcal{C}_b}.
\end{equation*}
We thus need to estimate on each block $1 \leq b \leq C$, each site $1 \leq i \leq n$ and each specie $j\in\mathcal{C}_b$
\begin{equation} 
\label{eqn:condMoments-CL}
	\esp[\btheta^{(b,h)}]{Z^{(b)}_{ij} \mid Y^{(b)}_{ij}},
	\quad 
	\esp[\btheta^{(b, h)}]{\exp(Z^{(b)}_{ij}) \mid Y^{(b)}_{ij}} 
	\quad \text{and} \quad
	\esp[\btheta^{(b, h)}]{{\bZ^{(b)}_i}^\top {\bSigma^{(b)}}^{-1} \bZ^{(b)}_i \mid \bY^{(b)}_i}.
\end{equation}
In the sequel, while $ \bZ^{(b)}$ denotes the $n\times k$ matrix of latent variables for the block $\mathcal{C}_b$, $Z^{(b)}_{ij}$ does not refer to an element of the $j$-th column of $\bZ^{(b)}$ but to an element of the column of $\bZ^{(b)}$ associated to specie $j$.

Conversely, the data block $\bY^{(b)}$ contributes only to the estimation of the elements $\btheta^{(b)}$. As a consequence, to be able to estimate each element of the covariance matrix $\bSigma$, we need to resort to overlapping blocks, so that each pair of species $(j, \ell)$ appears at least once in the same block. The construction of the blocks will be discussed in Section \ref{sec:blocks}.

\section{Importance Sampling within expectation-maximization algorithm \label{sec:ISEM}}

Monte-Carlo approximations of EM (MCEM) are a popular alternative to deterministic (\textit{e.g.}, variational) approximations to deal with intractable E-steps while preserving maximum likelihood estimator properties.
However, efficiently sampling from the PLN conditional $p_{\btheta}(\bZ \mid \bY)$ is not straightforward.
While the MCEM method allows extensive choices of sampling routines, we focus on importance sampling 
because, in addition to handling the intractability of the conditional distribution, it can benefit from existing variational approaches recalled in Section \ref{sec:VEM}, and is a readily parallel method that requires no warm-up phase.

\subsection{Maximum likelihood inference using importance sampling \label{sec:ML-ISEM}}

In the framework of the PLN model \eqref{eqn:pln}, because the sites $1 \leq i \leq n$ are supposed to be independent, the function $Q$ decomposes into
\begin{equation*}
Q(\btheta \mid \btheta^{(h)}) = \sum_{i = 1}^n \esp[\btheta^{(h)}]{\log p_{\btheta}(\bY_i, \bZ_i) \mid \bY_i}.
\end{equation*}
For each site $1 \leq i \leq  n$, the core of the problem is to estimate, for some measurable function $f$, quantities of the form
\begin{equation*}
	\esp[\btheta^{(h)}]{f(\bZ_i) \mid \bY_i} = \int_{\mathbb{R}^p} f(\bz) p_{\btheta^{(h)}}(\bz \mid \bY_i) \d\bz.
\end{equation*}
Importance sampling estimates such an expectation
by approximating {intractable} $p_{\btheta^{(h)}}(\cdot \mid \bY_i)$  with a random probability measure based on weighted samples from a probability density function $q^{(h)}_i$, called proposal distribution, {whose support contains the support of $p_{\btheta^{(h)}}(\cdot \mid \bY_i)$}.
The importance weights thus write as
\begin{equation*}
	\frac{p_{\btheta^{(h)}}(\cdot \mid \bY_i)}{q^{(h)}_i} = \frac{\rho^{(h)}_i}{\displaystyle\int_{\mathbb{R}^d} \rho^{(h)}_i(\bz) q^{(h)}_i(\bz) \d\bz}
	\quad\text{where}\quad
	\rho^{(h)}_i = \frac{p_{\btheta^{(h)}}(\bY_i, \cdot)}{q^{(h)}_i}.
\end{equation*}
With this change of integrating measure,
the related self-normalized importance sampling estimator using $N\in\mathbb{N}^*$ independent samples $(\bV_{i1}, \ldots, \bV_{iN})$ from $q^{(h)}_i$ is defined as
\begin{equation}
\label{eqn:wis}
	\is[q^{(h)}_i]{f(\bZ_i)}
	 \triangleq \sum_{r = 1}^N w^{(h)}_{ir} f(\bV_{ir}),
	\quad\text{with}\quad
	w^{(h)}_{ir} = \frac{\rho^{(h)}_i(\bV_{ir})}{\sum_{s = 1}^N \rho^{(h)}_i(\bV_{is})}.
\end{equation}
The latter provides a consistent estimator that converges at a rate $\sqrt{N}$ and is asymptotically unbiased. Self-normalized estimates are however biased for fixed $N$. 
While it may be neglected at first since it decreases faster than the variance with $N$, it is possible to account for it using the method developed by \cite{middleton19}.

\paragraph*{Estimators for the maximum likelihood inference in the PLN model} 
Self-normalized importance sampling estimators can therefore be applied to estimating the conditional moments \eqref{eqn:condMoments}, or more practically, their gradient counterparts involved in the M-step. 
Appendix \ref{app:updt-em} shows how to perform a stochastic gradient scheme to achieve the update for the regression coefficients $\bBeta$ and provide an update for the covariance matrix $\bSigma$ from the same weighted sample of a proposal distribution $q_i^{(h)}$. 
These updates solely require computing for each site $i$
\begin{equation}
\label{eqn:updt-is}
\left\lbrace 
\is[q^{(h)}_i]{
\exp(Z_{ij})
}
\right\rbrace_{1\leq j \leq p}
\quad\text{and}\quad
\is[q^{(h)}_i]{\bZ_i\bZ_i^\top}.
\end{equation}
The whole inference procedure is referred to as importance sampling EM (ISEM) and is described in Algorithm \ref{algo:is-em}.

As for the variance of the estimators, the PLN model forms an example where we cannot either exactly compute the conditional expectation \eqref{eqn:louis-1}. 
Nonetheless, stemming from this identity, we can derive a plug-in estimator for the score function $\nabla_{\btheta} \log p_{\btheta}(\bY_i)$ by recycling the particles and the weights from the importance sampling scheme of the last iteration of the ISEM.
Namely, the estimator \eqref{eqn:obs-fim} becomes
\begin{equation} \label{eqn:fisherISML}
\widehat{\bI}_n(\btheta) = \frac{1}{n}\sum_{i = 1}^{n} 
\is[q^{(h)}_i]{\nabla_{\btheta} \log p_{\btheta}\left(\bY_i, \bZ_i\right)}\is[q^{(h)}_i]{\nabla_{\btheta} \log p_{\btheta}\left(\bY_i, \bZ_i\right)}^\top.
\end{equation}
We shall remark that, to compute the plug-in estimator we do not have to store the particles and the weights associated with each site. Indeed, for each site $i$, the plug-in estimator is based on the same self-normalized importance sampling estimators \eqref{eqn:updt-is}  used in the M-step (see Appendix \ref{app:updt-em}).

\subsection{Maximum composite likelihood inference using importance sampling \label{sec:CL-ISEM}}

Importance sampling performances may quickly degrade when the number of species increases, making the inference difficult, if not impossible, for problems of the scale of the Barents Sea data.
The limitation of importance sampling arises from the difficulty of finding a proposal well-fitted to the, possibly, intricate target distribution.
Yet, controlling the discrepancy between the proposal and target distributions is essential to further ensure the computational efficiency of the method \citep{agapiou17, CD18}.
Algorithm \ref{algo:is-em} may therefore become inefficient 
to handle a few tens of variables, likewise in synecological datasets, and reducing the dimension of the space in which importance sampling is performed becomes paramount.

The benefit of Algorithm \ref{algo:CL-EM} is that, when the conditional moments in the CL-E-step are intractable, they can be estimated using importance sampling but with a sampling space of smaller dimension, \textit{i.e.}, of the size of the block. Specifically, we can consider a proposal distribution $q^{(b, h)}_i$ specific to each iteration $h$, block $\mathcal{C}_b$ and site $i$ and the self-normalized weights for a $N$-sample $(\bV_{i1}, \ldots, \bV_{iN})$ from $q^{(b, h)}_i$ writes as
\begin{equation}
\label{eqn:wis-composite}
w^{(b, h)}_{ir} = \frac{\rho^{(b, h)}_i(\bV_{ir})}{\sum_{s = 1}^N \rho^{(b, h)}_i(\bV_{is})}, \quad\text{where}\quad \rho_i^{(b, h)} = \frac{p_{\btheta^{(b, h)}}(\bY_i^{(b)}, \cdot)}{q^{(b, h)}_i}.
\end{equation}

\paragraph*{Estimators for the maximum composite likelihood inference in the PLN model} 
As in the case of the regular likelihood, importance sampling is used to estimate gradient counterparts of the moments \eqref{eqn:condMoments-CL}. 
Appendix \ref{app:updt-cem} states that, unlike the regular framework, we do not have access to a direct importance sampling estimator for $\bSigma$, but we can still perform a stochastic gradient scheme to achieve the updates for both parameters that relies, for each site $i$, each block $b$ and each specie $j$, on
\begin{equation*}
\sum_{i=1}^{n} \is[q^{(b, h)}_i]{\bZ_i^{(b)}{\bZ_i^{(b)}}^{\top}}
\quad\text{and}\quad
\sum_{b\in\mathcal{C}(j)} \is[q^{(b, h)}_i]{\exp\left(Z^{(b)}_{ij}\right)}.
\end{equation*}
The whole inference procedure is referred to as composite ISEM and is described in Algorithm \ref{algo:composite-is-em}.

The estimation of the Godambe information matrix \eqref{eqn:godambe} relies on the following Monte Carlo estimators
\begin{equation}
\label{eqn:H-J-ISMCL}
\widehat{\bJ}_n(\btheta) = \frac{1}{n}\sum_{i = 1}^n \widehat{\bS}_i \widehat{\bS}_i^{\top} 
\quad\text{and}\quad 
\widehat{\bH}_n(\btheta) = \frac{1}{n}\sum_{i = 1}^n\sum_{b = 1}^C \lambda_b \widehat{\bS}^{(b)}_i {\widehat{\bS}_i^{(b)\top}},
\end{equation}
where
\begin{equation*}
\widehat{\bS}_i^{(b)} = \is[q^{(h)}_i]{\nabla \log p_{\btheta}\left(\bY^{(b)}_i, \bZ^{(b)}_i\right)}^{\blacksquare}
\quad\text{and}\quad
\widehat{\bS}_i = \sum_{b = 1}^C\lambda_b \widehat{\bS}_i^{(b)},
\end{equation*}
where $\bS^{\blacksquare}$ stands for a vector $\bS$ of length $d \times k + k(k+1)/2$ that has been embedded into a vector $\mathbf{T} = (T_j)$ of length $d \times p + p(p +1)/2$ with $T_{j}$ taking value 0 if the parameter $\btheta_j$ is not involved in the block $\mathcal{C}_b$ and being equal to the coefficient of $\bS$ related to the parameter $\btheta_j$ otherwise.
From a numerical perspective, the computation of the former estimators is somewhat less straightforward than those used for the Fisher information matrix. These matrices do not depend on precisely the same statistics as those involved in the M-step, owing to differences in the cross-products utilized. We need to store the statistics specific to each block and each site.

\subsection{Importance proposal distribution \label{sec:Proposal-IS}}

The choice of the proposal distribution $q_i^{(h)}$ or $q_i^{(b,h)}$ is paramount to get reliable estimates with a reasonable simulation cost. 
Finding an appropriate proposal poses a challenge that can then be treated by adaptive sequential methods \citep[\textit{e.g.,}][]{cornuet2012adaptive, DDP21, KP22, DDR23}, but the latter remain computationally cumbersome to be plugged into an MCEM scheme as the target changes at each iteration.

Due to the independence between the sites, a different proposal distribution is set for each site $i$.
For a given site, we choose here to build the initial proposal distribution upon the variational approximation from \cite{CMR18, CMR19} and subsequently adapt it at each iteration in order to account for some dependencies between the species in the proposal distribution. 
Our approach relies on the following result (proof is given in Appendix \ref{proof:weight-norm}) that gives conditions, when the proposal distribution $q_i^{(h)}$ is Gaussian, to achieve non-normalized weights $\rho^{(h)}_i(\bV_{i})$ that have finite variance or are bounded. 

\begin{proposition} \label{prop:weight-norm}
	Under the PLN model \eqref{eqn:pln}, given $1 \leq i \leq n$ and a parameter value $\btheta = (\bBeta, \bSigma)$, letting $\bm \in \mathbb{R}^p$ and $\bS$ be a symmetric positive definite matrix and denoting $\varphi\left(\cdot \,;\, \bm, \bS\right)$ the density function on $\mathbb{R}^p$ of the multivariate normal distribution $\mathcal{N}(\bm, \bS)$, it holds that:
	\begin{enumerate}
		\item If $2{\bSigma}^{-1} - \bS^{-1}$ is positive definite, then 
		\label{item:is-finite-var}
		\begin{equation*}
			\int_{\mathbb{R}^p} \frac{p_{\btheta}(\bY_i, \bv)^2}{\varphi(\bv\,;\,\bm, \bS)}\d\bv  < \infty;
		\end{equation*}
		\item If ${\bSigma}^{-1} - \bS^{-1}$ is positive definite, then we also have
		\label{item:is-bounded-w}
		\begin{equation*}
			\sup_{\bv \in\mathbb{R}^p} \frac{p_{\btheta}(\bY_i, \bv)}{\varphi(\bv\,;\, \bm, \bS)}  < \infty.
		\end{equation*}
	\end{enumerate}
\end{proposition}

\paragraph*{Vanilla Gaussian proposal}
We might consider the Gaussian that best fits the conditional $p_{\btheta^{(h-1)}}(\cdot \mid \bY_i)$.
The proposal $q_i^{(h)}$ is then first set to the Gaussian approximation from variational inference (when $h=0$), then to a Gaussian matching the first two moments of  $p_{\btheta^{(h-1)}}(\cdot \mid \bY_i)$.
However, this choice is marred by a major drawback. 
The non-normalized weight $\rho^{(h)}_i(\bV_{i})$ still involves a dominant quadratic term related to the marginal variance $\bSigma^{(h)} = (\sigma_{jk}^{(h)})$ (see Equation \eqref{eqn:weight-norm} in Appendix \ref{proof:weight-norm}). 
Using a proposal based on the conditional variance yields a distribution that is more concentrated than the marginal of the latent variable, leading to unbounded weight in the tails. 
Moreover, the finite variance condition from Proposition \ref{prop:weight-norm} does not hold in all generality. 
Such a proposal distribution thereby provides a fair idea of where to sample, but may nonetheless lead to a zero effective sample size, that is, none of the simulated variables are influential in the estimation.

Conversely, using a Gaussian proposal scaled to the marginal variance of $\bZ_i$ ensures, at least, finite variance for the weights but results in samples that are too spread out compared to samples from the conditional. This results in suboptimal target exploration and low effective sample size.

\paragraph*{ISEM algorithm for the PLN model}
In what follows, we thus set $q^{(h)}_i$ to a mixture distribution that strikes a balance between the benefits offered by both aforementioned ones, namely, for $\alpha\in[0,1)$ 
\begin{equation}
\label{eqn:mixt}
q^{(h)}_i = \alpha \varphi\left(\cdot \,;\, \bm^{(h)}_i, \bS^{(h)}_i\right) + (1 - \alpha) \varphi\left(\cdot \,;\,\bm^{(h)}_i, \bSigma^{(h)}\right),
\end{equation}
where $(\bm^{(0)}_i, \bS^{(0)}_i)$ and $\bSigma^{(0)}$ are respectively the variational parameter $(\bm_i, \bS_i)$ and the variational estimation of $\bSigma$, and for $h\geq 1$, $\bm^{(h)}_i$ and $\bS^{(h)}_i$ are Monte Carlo estimates of the mean and variance of the conditional distribution $p_{\btheta^{(h-1)}}(\cdot \mid \bY_i)$. The following proposition states that the importance weights associated with the mixture inherit the same properties as the non-normalized weight related to $\varphi(\cdot \,;\, m, \bSigma^{(h)})$. The proof follows directly from Proposition \ref{prop:weight-norm}  and is given in Appendix  \ref{proof:weight-norm}.

\begin{proposition}
\label{prop:weight-mixt}
Under the same condition as Proposition \ref{prop:weight-norm}, with $\btheta^{(h)} = (\bBeta^{(h)}, \bSigma^{(h)})$, for all $\alpha \in[0,1)$, the non-normalized weights associated to the proposal distribution \eqref{eqn:mixt} satisfy
\begin{equation*}
\esp[q_i^{(h)}]{\rho_i^{(h)}(\bV)^2} = \int_{\mathbb{R}^p} \frac{p_{\btheta^{(h)}}(\bY_i, \bv)^2}{q_i^{(h)}(\bv)}\d\bv  < \infty.
\end{equation*}
\end{proposition}
The mixture proportion $\alpha$ controls how close we aim to be to the optimal Gaussian proxy of the conditional, while the second component is solely dedicated to ensuring finite variance of the non-normalized weights in Proposition \ref{prop:weight-mixt}. 
Algorithm \ref{algo:is-em} summarizes our ISEM algorithm for the inference of the PLN model \eqref{eqn:pln} based on the mixture proposal \ref{eqn:mixt}.

We refer the reader to Appendix \ref{sec:comp-ais} for a comparison of the computational performances of such a proposal with an M-PMC adaptive version.

\begin{algorithm2e}[!t]
\label{algo:is-em}
\caption{ISEM for PLN}
\KwIn{number of iterations $n_{\rm iter}$, number of draws $N$, mixture proportion $\alpha$, variational parameter $\{(\bm^{(0)}_i, \bS^{(0)}_i)\}_{1 \leq i  \leq n}$, variational estimation $\bSigma^{(0)}$ of $\bSigma$.}
\For{$h = 0$ \KwTo $n_{\rm iter} - 1$}{
	\For{$i = 1$ \KwTo $n$}{
		{\bf sample} $(\bV_{i1}, \ldots, \bV_{iN})$ from $q^{(h)}_i \sim \alpha \mathcal{N}\left(\bm^{(h)}_i, \bS^{(h)}_i\right) + (1 - \alpha)\mathcal{N}\left(\bm^{(h)}_i, \bSigma^{(h)}\right)$\;
		{\bf compute} $\left(w^{(h)}_{i1}, \ldots, w^{(h)}_{iN}\right)$ according to \eqref{eqn:wis}\;
		{\bf compute} $\is[q^{(h)}_i]{\bZ_i}$, 
		$\is[q^{(h)}_i]{\bZ_i\bZ_i^{\top}}$ 
		and $\left\lbrace \is[q^{(h)}_i]{\exp(Z_{ij})}\right\rbrace_{1\leq j \leq p}$\;
		{\bf set} 
		\[
		\bm^{(h+1)}_i = \is[q^{(h)}_i]{\bZ_i} \quad\text{and}\quad \bS^{(h+1)}_i = \is[q^{(h)}_i]{\bZ_i\bZ_i^{\top}} - \bm^{(h+1)}_i{\bm^{(h+1)}_i}^\top\,;
		\]
	}
	\tcc{See Appendix \ref{app:updt-em} for update formulas}
	{\bf set} $\bSigma^{(h + 1)} = \displaystyle\frac{1}{n}\sum_{i = 1}^n \is[q^{(h)}_i]{\bZ_i\bZ_i^{\top}}$\;
	\For{$j = 1$ \KwTo $p$}{ 
		{\bf set} $\bbeta_j^{(h + 1)}$ with a gradient scheme based on $\left\lbrace \is[q^{(h)}_i]{\exp(Z_{ij})}\right\rbrace_{1\leq i \leq n}$\;
	}
}
\end{algorithm2e}

\begin{algorithm2e}[!ht]
\label{algo:composite-is-em}
\caption{Composite ISEM for PLN}
\KwIn{number of iterations $n_{\rm iter}$, number of draws $N$, mixture proportion $\alpha$, variational parameter on each block $\{(\bm^{(b, 0)}_i, \bS^{(b, 0)}_i)\}_{1 \leq i  \leq n}$, variational estimation $\bSigma^{(0)}$ of $\bSigma$.}
\For{$h = 0$ \KwTo $n_{\rm iter} - 1$}{
	\For{$b = 1$ \KwTo $C$}{
		{\bf set} $\bSigma^{(b, h)} = [\sigma^{(h)}_{jk}]_{(j,k)\in\mathcal{C}_b}$\;
		\For{$i = 1$ \KwTo $n$}{
			{\bf sample} $(\bV^{(b)}_{i1}, \ldots, \bV^{(b)}_{iN})$ 
			from $q^{(b, h)}_i \sim \alpha \mathcal{N}\left(\bm^{(b, h)}_i, \bS^{(b, h)}_i\right) + (1 - \alpha)\mathcal{N}\left(\bm^{(b, h)}_i, \bSigma^{(b, h)}\right)$\;
			{\bf compute} $\left(w^{(b, h)}_{i1}, \ldots, w^{(b, h)}_{iN}\right)$ according to \eqref{eqn:wis-composite}\;
			{\bf compute} $\is[q^{(b, h)}_i]{\bZ^{(b)}_i}$, 
			$\is[q^{(b, h)}_i]{\bZ^{(b)}_i{\bZ^{(b)}_i}^{\top}}$ 
			and $\left\lbrace \is[q^{(b, h)}_i]{\exp(Z^{(b)}_{ij})}\right\rbrace_{j \in \mathcal{C}_b}$\;
			{\bf set} 
			\[
			\bm^{(b, h+1)}_i = \is[q^{(b, h)}_i]{\bZ^{(b)}_i} \quad\text{and}\quad \bS^{(b, h+1)}_i = \is[q^{(b, h)}_i]{\bZ^{(b)}_i{\bZ^{(b)}_i}^{\top}} - \bm^{(b, h+1)}_i{\bm^{(b, h+1)}_i}^\top\,;
			\]
		}
	}
	\tcc{See Appendix \ref{app:updt-cem} for update formulas}
	{\bf set} $\bSigma^{(h + 1)}$ with a gradient scheme based on $\left\lbrace \sum_{i=1}^{n} \is[q^{(b, h)}_i]{\bZ_i^{(b)}{\bZ_i^{(b)}}^{\top}}\right\rbrace_{1\leq b \leq C}$\;
	\For{$j = 1$ \KwTo $p$}{
		{\bf set} $\bbeta_j^{(h + 1)}$ with a gradient scheme based on $\left\lbrace \sum_{b\in\mathcal{C}(j)} \is[q^{(b, h)}_i]{\exp\left(Z^{(b)}_{ij}\right)}\right\rbrace_{1\leq i \leq n}$\;
	}
}
\end{algorithm2e}

\paragraph*{Composite ISEM algorithm for the PLN model}  
The proposal distribution \eqref{eqn:mixt} can easily be extended to the composite likelihood framework, where the aim is to have a mixture on each block. Namely, the proposal distribution for block $\mathcal{C}_b$ at iteration $h$ is
\begin{equation*}
q^{(b, h)}_i = \alpha \varphi\left(\cdot \,;\, \bm^{(b, h)}_i, \bS^{(b, h)}_i\right) + (1 - \alpha) \varphi\left(\cdot \,;\,\bm^{(b, h)}_i, \bSigma^{(b, h)}\right),
\quad\alpha\in(0,1),
\end{equation*}
where the parameter corresponds to those from \eqref{eqn:mixt} reduced to block $\mathcal{C}_b$, \textit{i.e.},
\begin{equation*}
\{(\bm^{(b, 0)}_i, \bS^{(b, 0)}_i)\}_{1 \leq i  \leq n} = \{([\bm_{i,j}], [\bS_{i,j}])\}_{1 \leq i  \leq n, j\in\mathcal{C}_b},
\qquad
\bSigma^{(b, 0)} = [\sigma^{(0)}_{jk}]_{(j,k)\in\mathcal{C}_b},
\end{equation*}
the updates $\bm^{(b, h)}_i$ and $\bS^{(b, h)}_i$ are the estimated mean and variance of the conditional distribution $p_{\btheta^{(b, h-1)}}(\cdot \mid \bY^{(b)}_i)$, while $\bSigma^{(b, h)}$ is the current estimate of $\bSigma^{(b)}$, that is $\bSigma^{(b, h)} = [\sigma^{(h)}_{jk}]_{(j, k) \in \mathcal{C}_b}$. The whole inference procedure for the PLN model and this peculiar choice of proposal is summarized in Algorithm \ref{algo:composite-is-em}.

\section{Illustrations}  \label{sec:illus}

In this section, we first discuss the construction of the blocks in view of composite likelihood inference. Then, we compare and assess the performances of the different inference algorithms previously proposed on synthetic datasets. We also present a comparison with the original VEM algorithm.

\subsection{Determining the blocks for composite likelihood inference} \label{sec:blocks}

We now discuss the definition of the blocks $\mathcal{C}_1 \dots \mathcal{C}_C$. 
Observe that, as opposed to the case of spatial data, the block can be constructed according to the proximity between the observation sites \citep[see, e.g.][]{ESR14, BeG15}, no additional information about the species is available here to help us in this task.
In this paper, we only considered that species are spread out into blocks with constant size $k \leq p$. 
Obviously, for a given block size $k$, using a small number of blocks $C$ alleviates the computational burden. 

To get an estimate of each entry of the covariance matrix $\bSigma$, it is sufficient that each pair of distinct species $(j, j')$, $1 \leq j < j' \leq p$, appears at least once in a same block $\mathcal{C}_b$, $1 \leq b \leq C$. 
Hence, it is unnecessary to explore all possible combinations of blocks with size $k$, so $C \leq {{p}\choose{k}}$. 
On the other hand, because there are $p(p-1)/2$ pairs of species and because each block contains $k(k-1)/2$ pairs (giving a total of $Ck(k-1)/2$ pairs), we need that $C \geq p(p-1)/[k(k-1)]$. 
Remark that $k=2$ is a trivial case as each block contributes to estimate one single covariance parameter, so $C = p(p-1)/2$.

\begin{figure}[!t]
  \begin{center}
    \includegraphics[width=.5\textwidth]{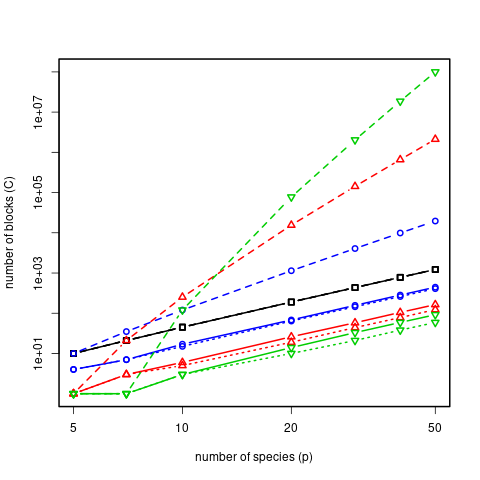}
  \end{center}
  \caption{
  Number of blocks $C$ as a function of the number of species $p$ (in log-log-scale) for blocks of size $k=2$ (black squares $\blacksquare$), $k=3$ (blue circles \textcolor{blue}{$\medcircle$}), $k=5$ (red triangles up \textcolor{red}{$\triangle$}) and $k=7$ (green triangles down \textcolor{green}{$\triangledown$}). Solid line: number of blocks actually used, dashed line: upper bound ${{p}\choose{k}}$, dotted line: lower bound $p(p-1)/[k(k-1)]$.}
  \label{fig:blockNb}
\end{figure}

Spreading the $p$ species into $k$ blocks is equivalent to build an incomplete block design in terms of design of experiments. 
In the general case, finding an incomplete block design with a minimal number of blocks is a challenging combinatorial task. 
We conceived a greedy stochastic algorithm to build such a design.
Figure \ref{fig:blockNb} gives the number of blocks returned by this algorithm for the various configurations of the simulation study. 

We observe that the upper bound ${{p}\choose{k}}$ is much too pessimistic, as compared to the obtained number of blocks $C$.
Our algorithm finds a number of blocks close to the lower bound $p(p-1)/[k(k-1)]$. 
Furthermore, we note a strong dependence of $C$ on $k$. 
For instance, taking $k=5$ yields a smaller number of blocks than taking $k=2$. 
Consequently, we expect better computational efficiency for the composite ISEM method (see Algorithm \ref{algo:composite-is-em}) when using blocks of size 5 rather than blocks of size 2. 
Due to $C$ increasing as $p^2$ for $k = 2$, the method is manageable for the Barents Sea dataset but would become inefficient for datasets with $p > 30$.

\subsection{Simulation study} \label{sec:simuls}

\newcommand{\lagSim}{50} \newcommand{\nISsim}{200} \newcommand{\nIterSim}{1000}
\newcommand{\simulParms}{-nIter\nIterSim-lag\lagSim-nIS\nISsim}

\subsubsection{Simulation design} 

To mimic typical datasets encountered in community ecology or biogeography, we fixed the number of sites to $n = 100$ and the number of covariates to $d=3$ (that is, one intercept and two covariates) and made the number of species vary from $p=5$ to $p=50$. 
The offset term was set to zero. For each dimension $p$, we fixed the $n \times d$ matrix of covariates $\bX$, the $d \times p$ matrix of regression coefficients $\bBeta$, and the $p \times p$ covariance matrix $\bSigma$. 
We sampled $M = 100$ count matrices $\bY^m$ ($1 \leq m \leq M$) 
according to the PLN model \eqref{eqn:pln}, each with dimension $n \times p$.

\paragraph*{Estimation algorithms} 
When the number of species $p$ is lower than 10, we carried out maximum likelihood inference using our ISEM method (see Algorithm \ref{algo:is-em}), and referred to as ``full likelihood'' (FL). 
For each simulated dataset $\bY^m$, we obtained the parameter estimates $\widehat{\bBeta}^m$ and $\widehat{\bSigma}^m$, along with their respective estimated asymptotic variance related to the estimated Fisher information matrix \eqref{eqn:fisherISML}.

For each configuration of the number of species and each simulated dataset, we also performed composite likelihood inference using our composite ISEM method (see Algorithm \ref{algo:composite-is-em}) with blocks of size $k = 2, 3, 5$ and $7$. 
It is further referred to as ``composite likelihood'' or ``CL$k$''. 
For each simulation setting, we computed the parameter estimates $\widehat{\bBeta}^m$ and $\widehat{\bSigma}^m$, and their respective estimated asymptotic variance, which was achieved with the estimates \eqref{eqn:H-J-ISMCL} of the Godambe matrix.

For all ISEM algorithms, we used a linearly increasing number of particles along the iterations: at iteration $h$, we used $N^{(h)} = h N^{(0)}$ particles. 
The initial number of particles was set to  $N^{(0)} = 50, 100$, and $200$. 
We did not observe any significant difference, with the exception of the situations (not shown) that are actually doomed to fail with all considered numbers of particles due to the aforementioned limitation of importance sampling when the dimension $p$ or $k$ grows.
We report hereafter the results obtained with $N^{(0)} = \nISsim$ initial particles. 
We set a maximum of {\nIterSim} iterations for each algorithm and used a lag of {\lagSim} steps for the stopping criterion. 
Regarding the mixture proposal distribution defined in Equation \eqref{eqn:mixt}, we used the mixture proportion $\alpha = 0.9$.

We also ran the variational EM algorithm implemented in the {\tt PLNmodels} R package. As mentioned in Section \ref{sec:Proposal-IS}, the resulting variational parameters are used as initial values for the ISEM algorithm  $\bm^{(0)}_i$ and $\bS^{(0)}_i$. 
Furthermore, as stated in Section \ref{sec:VEM}, two variance estimates for $\bBeta$ (but not for $\bSigma$) are proposed as outputs of the VEM algorithm: one based on the variational lower bound of the log-likelihood (denoted ``VEM'' in the sequel) and the other based on jackknife (denoted ``JK'' in the sequel). 
The two methods will be considered as two different algorithms, because, although they use the same estimation algorithm, they differ in terms of variance estimation.

\paragraph*{Normality of the estimators} 
The objective of this study is to assess the validity of the model parameters inference, especially in the ability to provide
tests and/or confidence intervals achieving the nominal level, which are not provided by variational inference. 
To this end, for each algorithm under consideration and each, say, regression parameter $\beta_{\ell j}$, we examined the standardized estimates
\begin{equation} \label{eq:betaTilde}
\tbeta_{\ell j} = \left(\widehat{\beta}_{\ell j} - \beta_{\ell j} \right) 
\left/ \sqrt{\widehat{\Var}\left[\widehat{\beta}_{\ell j}\right]} \right.\,,
\end{equation}
where $\beta_{\ell j}$ stands for the true value, $\widehat{\beta}_{\ell j}$ for its estimate provided by the algorithm and $\widehat{\Var}[\widehat{\beta}_{\ell j}]$ for the estimated variance of $\widehat{\beta}_{\ell j}$. 
According to the M-estimator theory, for a given set of parameters and a given ISEM algorithm, the distribution of the $\tbeta_{\ell j}^{(m)}$ across simulations $m = 1, \dots, M$ should be close to a standard normal. 

\subsubsection{Simulation results}

\paragraph*{Fit to the standard normal} 
We first used the Kolmogorov-Smirnov test to assess this normality and reported the resulting $p$-value.
Small $p$-values reveal a deviation from normality.
Figure \ref{fig:KSpval} gives the distributions of the $p$-values from the Kolmogorov–Smirnov test associated with the $p \times d$ regression parameters $\beta_{j\ell}$, obtained using the variational solutions VEM and JK, as well as the FL and CL$k$ algorithms, for $k = 2, 3, 5$, and $7$.
This figure demonstrates that the normality hypothesis (that is, the fit to the standard Gaussian $\mathcal{N}(0, 1)$) is not rejected for the CL$k$ algorithms, even for a large number of species ($p = 30$ or $50$). 
A departure from normality is however observed with full likelihood inference, for a moderate number of species ($p \simeq 10$). This illustrates the difficulty encountered by importance sampling to perform well, even for moderate values of $p$, while keeping a reasonable computational budget.
An in-depth analysis of the potential deviation from normality is presented in Appendix \ref{sec:betaDistCL}, paragraph ``Distribution of the estimates''. We observe notably that full likelihood estimates increasingly deviate as $p$ grows, primarily due to a systematic overestimation of the variance.

\begin{figure}[!t]
  \begin{center}
    \begin{tabular}{cccc}
      $p = 5$ & $p = 7$ & $p = 10$ & $p = 20$ \\
      \includegraphics[width=.225\textwidth, trim=10 10 25 25, clip=]{\dirsim/PvalKS-score-n100-d3-p5-parm1\simulParms} & 
      \includegraphics[width=.225\textwidth, trim=10 10 25 25, clip=]{\dirsim/PvalKS-score-n100-d3-p7-parm1\simulParms} & 
      \includegraphics[width=.225\textwidth, trim=10 10 25 25, clip=]{\dirsim/PvalKS-score-n100-d3-p10-parm1\simulParms} & 
      \includegraphics[width=.225\textwidth, trim=10 10 25 25, clip=]{\dirsim/PvalKS-score-n100-d3-p20-parm1\simulParms} \\
      \hline
      $p = 30$ & $p = 40$ & $p = 50$ \\
      \includegraphics[width=.225\textwidth, trim=10 10 25 25, clip=]{\dirsim/PvalKS-score-n100-d3-p30-parm1\simulParms} & 
      \includegraphics[width=.225\textwidth, trim=10 10 25 25, clip=]{\dirsim/PvalKS-score-n100-d3-p40-parm1\simulParms} & 
      \includegraphics[width=.225\textwidth, trim=10 10 25 25, clip=]{\dirsim/PvalKS-score-n100-d3-p50-parm1\simulParms} & 
    \end{tabular}
    \caption{
Distribution of the $p$-values from the Kolmogorov–Smirnov test applied to the distribution of the standardized estimates $\tbeta_{\ell j}$ over the $M = 100$ simulations, for each inference method: full likelihood (FL), composite likelihood with blocks of size $k$ (CL$k$), variational EM (VEM), and jackknife-based variational EM (JK). Each boxplot summarizes the $d \times p = 3p$ normalized coefficients $\tbeta_{\ell j}$. Dotted red lines: $\alpha = 5\%$ significance threshold after Bonferroni correction (\textit{i.e.}, $\alpha/(dp)$).
    }
    \label{fig:KSpval}
  \end{center}
\end{figure}

\paragraph*{Effect of the block size}
Clearly, each composite likelihood inference procedure produces different estimators, with different asymptotic variances. We investigated how the (estimated) asymptotic variance of the regression coefficient estimates varies with the block size $k$.
Figure \ref{fig:RatioVarBeta} displays the distribution (across coefficients) of the variance ratios, with the CL5 algorithm as an arbitrary reference to account for the intrinsic variability between coefficients. 
This figure illustrates that the variance exhibits a slight decline as $k$ increases, but consistently remains close to one, which suggests that the asymptotic variance is not too much affected by the choice of $k$.

\paragraph*{Computational time} 
Table \ref{tab:computTime} presents the mean computational times required by the CL$k$ algorithms for a range of species $p$, along with the number of iterations and the number of blocks. We first observe that the mean number of iterations is considerably smaller than the maximum number allowed (1000). As expected, the computational time mainly depends on the number of species $p$.
The influence of the block size $k$ is more intricate, as it has an impact on both the computation cost of each iteration and the total number of iterations. 
The sampling effort required for a single iteration depends on both the number of blocks and their sizes (you may have fewer blocks, but the cost of simulating one block may be greater). 
Table \ref{tab:unitaryComputTime} in Appendix \ref{sec:betaDistCL} gives the computational time for one block in one iteration. Furthermore, the number of iterations is governed by the geometry of the problem and is itself affected by both the number and size of the blocks.

\begin{table*}[!h]
    \caption{
Average computational performance over the $M = 100$ simulations, for $p = 10$, $30$, and $50$ species, and for each inference method: composite likelihood (CL$k$) with block sizes $k = 2, 3, 5$, and $7$, and variational EM (VEM). From left to right: the mean computational time (in seconds), the mean number of iterations, and the number of blocks used.
}
    \label{tab:computTime}
\tabcolsep=1pt
\begin{tabular*}{\textwidth}{@{\extracolsep{\fill}}lrrrrrrrrrrrrr@{\extracolsep{\fill}}}
\toprule%
& \multicolumn{5}{@{}c@{}}{Computational time (s)} & \multicolumn{4}{@{}c@{}}{Iterations} &  \multicolumn{4}{@{}c@{}}{Number of blocks} \\
\cmidrule{2-6}\cmidrule{7-10}\cmidrule{11-14}%
Algorithm & CL$2$ & CL$3$ & CL$5$ & CL$7$ & VEM & CL$2$ & CL$3$ & CL$5$ & CL$7$ & CL$2$ & CL$3$ & CL$5$ & CL$7$\\
\midrule
$p = 10$ & 44 & 42 & 216 & 149 & 10 & 124 & 115 & 107 & 101 & 45 & 17 & 6 & 3 \\ 
$p = 30$ & 4061 & 8936 & 592 & 3196 & 18 & 299 & 256 & 243 & 216 & 435 & 159 & 60 & 34 \\ 
$p = 50$ & -- & 10769 & 8744 & 4249 & 34 & -- & 522 & 443 & 123 & -- & 448 & 166 & 93 \\
\botrule
\end{tabular*}
\end{table*}

\paragraph*{Comparison with VEM estimates}
Our work is mainly motivated by the estimation of the (asymptotic) variance of the estimates, in order to provide confidence intervals or significance tests for the model's parameters. 
Natural competitors to our Monte Carlo solution are the variance estimates that arise from the variational paradigm (see Section \ref{sec:VEM}).
Table \ref{tab:computTime} shows that, despite the jackknife procedure requires to run the VEM algorithm $n=100$ times, the computational burden of VEM remains significantly lighter than that of the FL and CL$k$ algorithms we propose. Consequently, resorting to the jackknife estimator is appealing to carry out inference in an efficient manner.

Unfortunately, Figure \ref{fig:KSpval} shows the poor fit of both variational standardized estimates to the standard normal, regardless of the number of species, 
albeit the jackknife procedure is almost satisfying up to $p=10$ species. 
Figure \ref{fig:qqplot-BetaTildeP30} highlights the contrasting behaviour of the two estimates: the VEM approach clearly under-estimates the variance of the regression coefficients, whereas the jackknife tends to over-estimate it.
Hence, the VEM approach results in too narrow confidence intervals and an illusory powerful test, whereas the jackknife solution turns out to be conservative, resulting in a lack of power for hypothesis testing. 
These results underscore the importance and the necessity of moving away from variational variance estimates in favor of alternative, more accurate variance estimators, such as the one we propose.

Figure \ref{fig:qqplot-BetaTildeP30} further highlights the contrasting behavior of the two estimators: the VEM approach clearly underestimates the variance of the regression coefficients, while the jackknife tends to overestimate it.
As a result, the VEM approach produces overly narrow confidence intervals and misleadingly powerful tests, whereas the jackknife is conservative, leading to reduced power in hypothesis testing.
These findings underscore the need for alternative and more accurate variance estimators, such as the one we propose.

\paragraph*{Conclusion}
This numerical study points out that the CL class of algorithms represents a compelling class of methods for analysing the Barents Sea dataset or synecological data of a similar scale.
Indeed, the algorithms demonstrate consistent performances regardless of some calibration choices, where the level of approximation inherent to the variational solution leads to unreliable outputs, and the computational burden of the sampling method makes the maximum likelihood inference infeasible or dubious for a fixed budget.

\FloatBarrier

\section{Fish abundances in the Barents Sea}

\newcommand{\nIterEss}{50}
\newcommand{\nIterEx}{10000} \newcommand{\lagEx}{50} \newcommand{\nISex}{200}
\newcommand{\exampleParms}{-nIter\nIterEx-lag\lagEx-nIS\nISex}
\newcommand{\nIterSel}{\nIterEx} \newcommand{\lagSel}{20} \newcommand{\nISsel}{\nISex}
\newcommand{\selectParms}{-nIS\nISsel-nIter\nIterSel-lag\lagSel}

We now turn to the analysis of the Barents Sea dataset introduced in Section \ref{subsec:PLN}. 
We used the composite likelihood algorithm described in Section \ref{sec:CL} for different block sizes and the VEM algorithm from the {\tt PLNmodels} package. Following the observations of the previous section,
we started with $N^{(0)} = \nISex$ particles with a maximum number of iterations of {\nIterEx} and a lag of {\lagEx} steps for the stopping rule. 

\subsection{Reduced dataset: $p = 7$ species}

To begin with, we focus on the $p=7$ most abundant species of the dataset. 
This allows us to conduct maximum likelihood inference and to further compare the ISEM method (FL) with the composite ISEM (CL$k$) method described in Sections \ref{sec:ML-ISEM} and \ref{sec:CL-ISEM}.
For the composite likelihood, 
we followed the same strategy as in Section \ref{sec:simuls} with blocks of size $k=2, 3$ and $5$ ($k = 7$ corresponding to the FL framework). 
The number of iterations at convergence for each algorithm were FL = 2885, CL2 = 4172, CL3 = 4471 and CL5 = 3484, while the VEM algorithm from the {\tt PLNmodels} packages took 89 before convergence. 

\begin{figure}[t!]
  \begin{centering}
  \begin{tabular}{cc}
	Full likelihood & Composite likelihood ($k=2$) \\
    \includegraphics[width=.4\textwidth, trim=10 40 10 50, clip=]{\direx/BarentsLarge\exampleParms-ESS-em-iter\nIterEss} 
    &
    \includegraphics[width=.4\textwidth, trim=10 40 10 50, clip=]{\direx/BarentsLarge\exampleParms-ESS-cem2-iter\nIterEss} 
    \\
    \hline
    Composite likelihood ($k=3$) & Composite likelihood ($k=5$) \\
    \includegraphics[width=.4\textwidth, trim=10 40 10 50, clip=]{\direx/BarentsLarge\exampleParms-ESS-cem3-iter\nIterEss}
    &
    \includegraphics[width=.4\textwidth, trim=10 40 10 50, clip=]{\direx/BarentsLarge\exampleParms-ESS-cem5-iter\nIterEss}
  \end{tabular}
  \caption{
    Effective sample size (ESS) over the first {\nIterEss} iterations for the different inference methods on the reduced Barents Sea dataset ($p = 7$ species). From top to bottom: full likelihood (FL) and composite likelihood (CL$k$) with block sizes $k = 2, 3$, and $5$. One boxplot is shown per iteration, summarizing the ESS values across all blocks and all sites. Horizontal red dashed line: median ESS across the remaining iterations.
  }
  \label{fig:bar7ess}
  \end{centering}
\end{figure}

Figure \ref{fig:bar7ess} displays the evolution of the distribution of the standardized effective sampling sizes (ESS) across each site and each block.
The ESS is a real number in $[0, 1]$ defined for a site $i$ and a block $b$ as
\begin{equation*}
\mathrm{ESS}^{(b)}_i = \frac{1}{N}  \left(\sum_{r = 1}^N w^{(b)}_{ir}\right)^2 \left/ \left(\sum_{r = 1}^N \left(w^{(b)}_{ir}\right)^2\right)\right.\,.
\end{equation*}
It informs on the discrepancy between the weighted sample and the sample in which all realisations were equally weighted. A higher ESS indicates a better estimation as more particles are informative in the computation. In practice, one wants the ESS to be as close to 1 as possible.
We observe that the ESS rapidly (less than 50 iterations) reaches 80\% for most of the sites and blocks, meaning proposal distributions used across sites and blocks are well suited to our problem.

\begin{figure}[t!]
  \begin{centering}
    \begin{tabular}{ccc}
      Regression coefficients & Covariance parameters & Variance of the estimates  \\
\includegraphics[width=.3\textwidth, trim=15 40 25 50, clip=]{\direx/BarentsLarge\exampleParms-compBeta-em-all} &
    \includegraphics[width=.3\textwidth, trim=15 40 25 50, clip=]{\direx/BarentsLarge\exampleParms-compSigma-em-all} &
    \includegraphics[width=.3\textwidth, trim=15 40 25 50, clip=]{\direx/BarentsLarge\exampleParms-compVarBeta-em-all}
  \end{tabular}
  \caption{
Comparison of the estimates obtained using different inference methods on the Barents Sea reduced dataset ($p = 7$ species). Left: estimated regression coefficients $\widehat{\bBeta}=(\widehat{\beta}_{hj})$; center: estimated covariance parameters $\widehat{\bSigma} = (\widehat{\sigma}_{jk})$; right: estimated variances of the regression coefficients $\widehat{\beta}_{hj}$. $x$-axis: estimates from full likelihood inference (FL), $y$-axis:  estimates from the other methods: FL (gray asterisk [\textcolor{gray}{$*$}], used as the reference), composite likelihood (CL$k$) with blocks of size $k = 2$ (cyan diamond [\textcolor{cyan}{$\meddiamond$}]), $k = 3$ (green plus sign [\textcolor{green}{$+$}]), and $k = 5$ (red triangle [\textcolor{red}{$\triangle$}]), and variational EM (VEM, black circles [$\medcircle$]).
  }
  \label{fig:bar7parms}
  \end{centering}
\end{figure}

A comparison of the estimates obtained with the different algorithms is presented in Figure \ref{fig:bar7parms}. 
The estimates of the regression coefficients $\beta_{hj}$ are fairly consistent for both FL and CL$k$ algorithms, but depart from the variational estimates, which were used to initialize each of them. 
A similar observation holds for the covariance parameters $\sigma_{jk}$, with a greater diversity among the FL and CL$k$ algorithms. 
Importantly, the variance of the estimators of the regression coefficients $\widehat{\beta}_{hj}$ varies significantly between full-likelihood and composite-likelihood inference.
In particular, composite likelihood yields less variable estimates.
This suggests that, in the case of the PLN model, the latter proves to be more effective in detecting potential effects of the covariates.

\subsection{Full dataset: $p = 30$ species}

In our study of the complete dataset of \cite{FNA06}, we did not pursue the pairwise option. As observed in Section \ref{sec:simuls}, for $p = 30$, the choice $k = 2$ proves to be unnecessarily costly due to the block design without enhancing any improvement. We thereby focus on composite likelihood with solely $k=3, 5$ and 7.

\paragraph*{Comparison of the different algorithms} 
The left {panel} of Figure \ref{fig:BarentsAlgo} shows that the estimates of the regression coefficients $\beta_{\ell j}$ obtained with the three CL$k$ algorithms are all very close. 
Furthermore, the difference with the variational estimates is also quite small. 
A similar observation can be made about the covariance parameters $\sigma_{jk}$ (middle panel of Figure \ref{fig:BarentsAlgo}). 
These results point out that the estimates are robust to the choice of the block size $k$. 
It also corroborates the general observation that VEM provides accurate estimates of the parameters (without matching them with any uncertainty measure).

The right panel of Figure \ref{fig:BarentsAlgo} assesses the sampling efficiency of the CL5 algorithm, which stopped after 2571 iterations. 
The standardized effective sample size is, on the whole, higher than 80\% after less than 100 iterations, indicating efficient and reliable sampling with minimal weight variability for the various sites and blocks. 
As was the case for the reduced dataset, we observed equivalent efficiency for the CL3 and CL7 algorithms (not shown).

\begin{figure}[!t]
  \begin{center}
    \begin{tabular}{ccc}
      Regression coefficients & Covariance parameters & Effective sample size  \\
      \includegraphics[width=.3\textwidth, trim=10 10 25 50, clip=]{\direx/Barents\exampleParms-compBeta-cem5-all} & 
      \includegraphics[width=.3\textwidth, trim=10 10 25 50, clip=]{\direx/Barents\exampleParms-compSigma-cem5-all} & 
      \includegraphics[width=.3\textwidth, trim=10 10 25 50, clip=]{\direx/Barents\exampleParms-ESS-cem5}     
    \end{tabular}
    \caption{
Results on the Barents Sea full dataset ($p = 30$ species). Left: estimated regression coefficients $\widehat{\bBeta} = (\widehat{\beta}_{\ell j})$; center: estimated covariance parameters $\widehat{\bSigma} = (\widehat{\sigma}_{jk})$. $x$-axis: estimates obtained using the composite likelihood method $CLk$ with blocks of size $k = 5$ (CL5, red triangle [\textcolor{red}{$\triangle$}] = reference); $y$-axis:  estimates obtained using variational EM (VEM, black circles [$\medcircle$]), CL3 (green plus sign [\textcolor{green}{$+$}]), and CL7 (blue times sign [\textcolor{blue}{$\times$}]). Right: boxplot of the effective sample size across all sites and blocks as a function of the iteration number for the CL5 algorithm.
    }
    \label{fig:BarentsAlgo}
  \end{center}
\end{figure}

To further inquire into differences between the VEM and CL$k$ algorithms, we examined the proportion of VEM estimates covered by the composite likelihood confidence intervals. 
We observed that only 4 variational estimates of the regression coefficients $\beta_{\ell j}$ (out of $q \times p = 150$) felt outside the CL5 confidence intervals. 
Conversely, all variational estimates of the variance parameters $\sigma_{kj}$ were found to be within the CL5 confidence intervals. 
To this respect, only a very small fraction of the VEM estimates turned out to significantly differ from the corresponding CL5 estimates. 
Nonetheless, Table \ref{tab:BarentsVEM-CL5} shows that, when such discrepancies occur, they may lead to substantially different estimations of the regression parameters.

\begin{table}[!h]
\caption{
Comparison of variational EM (VEM) and composite likelihood with blocks of size $k = 5$ (CL5) estimates when the VEM estimate falls outside the 95\% confidence interval of the CL5 me\-thod, on the Barents Sea full dataset ($p = 30$ species).
}
\label{tab:BarentsVEM-CL5}
\begin{tabular*}{\columnwidth}{@{\extracolsep\fill}lrrrr@{\extracolsep\fill}}
\toprule
Regression parameter & $\beta_{1,1}$ & $\beta_{2,2}$ & $\beta_{2,2}$ & $\beta_{1,11}$ \\
\midrule
 VEM & -1.23 & 2.46 & 2.09 & -0.94 \\
 CL5 & -6.33 & 5.39 & 3.10 & -1.93 \\
\botrule
\end{tabular*}
\end{table}

\paragraph*{Model selection} 
As recalled at the end of Section \ref{sec:CL}, a BIC criterion can be associated with composite likelihood inference. 
Combining the four available covariates (latitude, longitude, temperature and depth), we fitted the $2^4$ possible models -- from Model 1: intercept only, Model 2: intercept plus latitude, Model 3: intercept plus longitude, up to Model 16: intercept plus the four covariates -- to the full Barents dataset. 
For each model, we estimated the dimension $\text{dim}(\btheta)$ defined in \eqref{eq:cl-bic} as $\widehat{\text{dim}}(\btheta) = \tr[\widehat{\bH}_n(\btheta) \widehat{\bG}_n(\btheta)^{-1}]$ where $\widehat{\bH}_n(\btheta)$ and $\widehat{\bJ}_n(\btheta)$ are given in \eqref{eqn:godambe} and $\widehat{\bG}_n(\btheta) = \widehat{\bH}_n(\btheta) \widehat{\bJ}_n(\btheta)^{-1} \widehat{\bH}_n(\btheta)$.

The left panel of Figure \ref{fig:BarentsBIC} displays the values of the composite log-likelihoods obtained with the CL5 and CL7 algorithms. 
In order to emphasize that the maximized composite likelihood
does not increase at the same speed, depending on the block size $k=5$ or $7$, we subtracted the composite log-likelihood for Model 1. 
This is consistent with the notion of adaptive dimension $\dim(\btheta)$ appearing in the BIC derived by \cite{GaS10} (see \eqref{eq:cl-bic}).

The center panel of Figure \ref{fig:BarentsBIC} shows that, although the estimate $\widehat{\text{dim}}(\btheta)$ of this adaptive dimension is a combination of the two Monte-Carlo estimates $\widehat{\bH}_n(\btheta)$ and $\widehat{\bJ}_n(\btheta)$, it remains almost constant across models with the same number of variables.

\begin{figure}[!t]
  \begin{center}
    \begin{tabular}{ccc}
      Composite likelihood & Dimension & BIC \\
      \includegraphics[width=.3\textwidth, trim=10 10 25 50, clip=]{\direx/Barents\selectParms-models-cl} & 
      \includegraphics[width=.3\textwidth, trim=10 10 25 50, clip=]{\direx/Barents\selectParms-models-penbic} & 
      \includegraphics[width=.3\textwidth, trim=10 10 25 50, clip=]{\direx/Barents\selectParms-models-bic}     
    \end{tabular}
    \caption{
    Model selection among the 16 possible models ($x$-axis) for the Barents Sea full dataset ($p = 30$ species), using the composite likelihood solution (CL$k$) with $k= 5$ and $7$. 
    Dashed vertical lines: limits between models involving $d=1, \dots 5$ covariates (including the intercept).
    Left: composite likelihood at the maximum composite likelihood estimates (subtracting the composite likelihood for Model 1); 
    Center: dimension of the parameter, as defined in Equation \eqref{eq:cl-bic}; 
    Right: BIC criterion (subtracting the composite likelihood for Model 1). CL5 algorithm: red triangle [\textcolor{red}{$\triangle$}], CL7: blue 'times' sign [\textcolor{blue}{$\times$}] (the best model within each dimension is circled for each algorithm).
    }
    \label{fig:BarentsBIC}
  \end{center}
\end{figure}

The final BIC criteria for the 16 models and the two algorithms CL5 and CL7 (still subtracting the composite log-likelihood of Model 1) is represented on the right panel of Figure \ref{fig:BarentsBIC}. 
When considering models with the same number of covariates, the BIC criterion does select the same model with
one covariate (latitude), two covariates (latitude + depth), and four covariates for the CL5 and CL7 algorithms. 
However, the model latitude + depth + temperature is selected with CL5, whereas CL7 yields the model latitude + depth + longitude. 
These differences are likely to result from the strong correlation between the temperature and both the latitude and the longitude that is observed in the dataset.

Overall, for both algorithms, the BIC criterion opts for the full model (Model 16), suggesting that all covariates should be kept in the model. 
This is consistent with the fact that, when considering a separate univariate PLN model for each of the $p=30$ species, each of the four covariates turns out to be significant for at least one of the species, even after a Bonferroni correction (not shown). We therefore focus on the interpretation of the full model in the next section.

It is noteworthy that this conclusion differs from that obtained using the variational approximation.
The variational inference of PLN models also comes with a so-called variational BIC criterion.
The latter consists of subtracting the regular BIC penalty $\log(n) D/2$ (where $D = pd + p(p+1)/2$ stands for the number of independent parameters) from the variational lower bound of the log-likelihood \citep[see][Equation (14)]{CMR18}. 
In the present case, the variational BIC opts for the same models as the CL5 algorithms within each dimension.
However, the overall best model turns out to include only the latitude and the depth. 
In this instance, the variational approximation also has consequences in terms of model selection.

\paragraph*{Interpretation} 
One of the main interests of the algorithm we propose here is to provide an accurate estimate of the variance of the parameter estimates of the PLN model. 
This enables the practitioner to determine ($i$) which environmental covariates have a significant effect on each species under study and ($ii$) which pairs of species display a correlation that does not result from environmental variations.

The results obtained are gathered in Figure \ref{fig:BarentsParm}. 
The left panel shows a contrasted pattern, which indicates that the different species are differentially affected by the different environmental conditions: for instance, deep waters favour species 30, whereas they disadvantage species 8.
The species displayed on top exhibit large estimated effect sizes $\beta_{\ell j}$, which indicates a strong sensitivity to changes, whereas species displayed at the bottom seem to be poorly affected, which suggests a broader ecological niche.

The center and right panels of Figure \ref{fig:BarentsParm} present the estimate of the latent covariance matrix $\bSigma$ (and of the associated correlation matrix $\text{cor}(\bSigma)$), where only significant terms are displayed. 
We observe a fairly sparse pattern, indicating that only a small fraction of species pairs (41 out of 435) have correlated abundance variations, once the environmental effects have been taken into account. 
Significant correlations indicate potential direct or indirect interactions between species.

\begin{figure}[!t]
  \begin{center}
    \begin{tabular}{ccc}
      Regression coefficients & Covariance matrix & Correlation matrix  \\
      \includegraphics[width=.3\textwidth, trim=10 10 25 25, clip=]{\direx/Barents\exampleParms-betaSignif-cem5} & 
      \includegraphics[width=.3\textwidth, trim=10 10 25 25, clip=]{\direx/Barents\exampleParms-sigmaSignif-cem5} &
      \includegraphics[width=.3\textwidth, trim=10 10 25 25, clip=]{\direx/Barents\exampleParms-corSigmaSignif-cem5}
    \end{tabular}
    \caption{
    Colormap of the parameter estimates for the Barents Sea full dataset ($p = 30$ species). Left: regression coefficients  $\widehat{\bBeta} = \widehat{\beta}_{\ell j}$; center: covariance parameters $\widehat{\bSigma} = (\widehat{\sigma}_{jk})$; right: corresponding correlations $\widehat{\rho}_{jk} = \widehat{\sigma}_{jk}/\sqrt{\widehat{\sigma}_{jj}\widehat{\sigma}_{kk}}$. Blank cells correspond to non-significant estimates, \textcolor{red}{red cells} (marked with '$+$') to positive estimates and \textcolor{blue}{blue cells} (marked with '$-$') to negative estimates.}
    \label{fig:BarentsParm}
  \end{center}
\end{figure}

\paragraph*{Comparison with variational inference} 
We can further assess the difference in terms of the interpretation of our algorithm compared with the variational EM algorithm
The simulation study from Section \ref{sec:simuls} outlined that the VEM approach underestimates the variance of the estimates, whereas the jackknife tends to overestimate them. 
This pattern is indeed observed with the Barents Sea dataset. The left panel of Figure \ref{fig:BarentsParmVEM} shows that the VEM yields test statistics with larger magnitude than those resulting from the composite inference, while the jackknife has the opposite behaviour.

We can examine the impact of using the VEM or the jackknife instead of the composite likelihood when interpreting the effects of the covariates. To this end, we display the VEM and jackknife counterparts of the left panel of Figure \ref{fig:BarentsParm} in the center and right panels of Figure \ref{fig:BarentsParmVEM}.
Obviously, the VEM (respectively, the jackknife) approach yields considerably more (respectively, less) significant effects than composite likelihood inference. 
More specifically, among the $p \times d = 150$ regression coefficients $\beta_{\ell j}$, the VEM approach identifies 79 significant effects (74 after Benjamini-Hochberg correction for multiple testing), the jackknife 27 (14 after correction), and the composite likelihood 37 (26 after correction). 
Given that composite likelihood inference is the only approach to enjoy (asymptotic) statistical guarantees, it is reasonable to conclude that variational inference combined with jackknife is likely to have limited power to detect significant effects, whereas the VEM approach results in numerous false positives.

\begin{figure}[!t]
  \begin{center}
    \begin{tabular}{ccc}
      Tests statistics & $\widehat{\bBeta}$: VEM & $\widehat{\bBeta}$: JK \\
      \includegraphics[width=.313\textwidth, trim=0 10 25 25, clip=]{\direx/Barents\exampleParms-betaStats-all-cem5} & 
      \includegraphics[width=.3\textwidth, trim=15 10 25 25, clip=]{\direx/Barents\exampleParms-betaSignif-vem} & 
      \includegraphics[width=.3\textwidth, trim=15 10 25 25, clip=]{\direx/Barents\exampleParms-betaSignif-jk}  
    \end{tabular}
    \caption{
    Significance of the regression coefficients for the Barents Sea full dataset ($p = 30$ species). 
    Left: comparison of the test statistics for all $\beta_{\ell j}$. $x$-axis = composite likelihood algorithm with blocks of size $k = 5$, $y$-axis = variational test statistics; blue triangles \textcolor{blue}{$\triangle$}: variational estimate of the variance (VEM), red circles \textcolor{red}{$\medcircle$}: jackknife estimate of the variance (JK). 
    Dashed lines = least-square regression lines.   
    Center: regression coefficients  $\widehat{\beta}_{\ell j}$ and significance according to the variational estimate of the variance. The species are displayed in the same order as in Figure \ref{fig:BarentsParm}
    Right: same coefficients and significance according to the jackknife estimate of the variance
    (same legend as Figure \ref{fig:BarentsParm}).
    }
    \label{fig:BarentsParmVEM}
  \end{center}
\end{figure}

\FloatBarrier

\section*{Code availability}
The proposed algorithms have all been implemented in R \citep{R} and C++, and will be included in the {\tt PLNmodels} package \citep{CMR21} soon. For the time being, the codes are available from the authors upon request.

\section*{Acknowledgements}
The authors thank Julien Chiquet (INRAE MIA-Paris-Saclay, France) and Mahendra Mariadassou (INRAE MaIAGE, France) for helpful discussions and advice.
The authors are grateful to the INRAE MIGALE bioinformatics facility (MIGALE, INRAE, 2020. Migale bioinformatics Facility, doi: 10.15454/1.5572390655343293E12) for providing computing resources. 
The first author has been partly funded by the European Union (ERC-2022-SYG-OCEAN-101071601). 
Views and opinions expressed are however those of the author only and do not necessarily reflect those of the European Union or the European Research Council Executive Agency. 
Neither the European Union nor the granting authority can be held responsible for them.

\bibliographystyle{abbrvnat}
\bibliography{Biblio}


\appendix
\section{Appendix}

\subsection{Proofs}

\subsubsection{Proof  of Proposition \ref{prop:CL-EM} \label{app:proof}}

The EM algorithm relies on the decomposition of the log-likelihood of the observed data $\bY$:
\begin{equation} \label{eqn:EM}
  \log p_{\btheta}(\bY) = \esp[\btheta]{\log p_{\btheta}(\bY, \bZ) \mid \bY} + \mathcal{H}_{\btheta}(\bZ \mid \bY),
\end{equation}
where $\mathcal{H}_{\btheta}(\bZ \mid \bY)$ stands for the conditional entropy of the unobserved variables $\bZ$ given the observed ones: 
\begin{equation*}
\mathcal{H}_{\btheta}(\bZ \mid \bY) = 
-\int_{\mathbb{R}^p} p_{\btheta}(\bz\mid\bY) \log p_{\btheta}(\bz\mid\bY)\d\bz,
\end{equation*}
and $\log p_{\btheta}(\bY, \bZ)$ is the so-called complete log-likelihood.
In the composite likelihood framework, the decomposition \eqref{eqn:EM} holds for each block of data $\bY^{(b)}$, so Equation \eqref{eqn:cl} can be rephrased as
\begin{equation} \label{eqn:clZ}
  \cl_{\btheta}(\bY) 
  = \sum_{b=1}^C \lambda_b \esp[\btheta]{\log p_{\btheta}(\bY^{(b)}, \bZ) \mid \bY^{(b)}} + \sum_{b=1}^C \lambda_b \mathcal{H}_{\btheta}(\bZ \mid \bY^{(b)}).
\end{equation}
Assuming $p_\btheta(\bZ\mid\bY^{(b)}) = p_\btheta(\bZ^{(b)}\mid\bY^{(b)})$, Equation \eqref{eqn:cl} becomes
\begin{equation} \label{eqn:clZb}
\cl_{\btheta}(\bY) 
= \sum_{b=1}^C \lambda_b \esp[\btheta]{\log p_{\btheta}(\bY^{(b)}, \bZ^{(b)}) \mid \bY^{(b)}}
+ \sum_{b=1}^C \lambda_b \mathcal{H}_{\btheta}(\bZ^{(b)} \mid \bY^{(b)}).
\end{equation}
The EM objective function is then
\begin{equation}
\label{eqn:QCISEM}
Q_{\cl}\left(\btheta\mid\btheta^{(h)}\right) = \sum_{b=1}^C \lambda_b \esp[\btheta]{\log p_{\btheta}(\bY^{(b)}, \bZ^{(b)}) \mid \bY^{(b)}}.
\end{equation}
As $\btheta^{(h+1)}$ is the maximizer of \eqref{eqn:QCISEM}, we have that
\begin{align*}
  0
  & \leq \sum_{b=1}^B \lambda_b \esp[\btheta^{(h)}]{\log p_{\btheta^{(h+1)}}(\bY^{(b)}, \bZ^{(b)}) \mid \bY^{(b)}} - \sum_{b=1}^B \lambda_b \esp[\btheta^{(h)}]{\log p_{\btheta^{(h)}}(\bY^{(b)}, \bZ^{(b)}) \mid \bY^{(b)}} \\
  & = \sum_{b=1}^B \lambda_b \esp[\btheta^{(h)}]{\log \frac{p_{\btheta^{(h+1)}}(\bY^{(b)}, \bZ^{(b)})}{p_{\btheta^{(h)}}(\bY^{(b)}, \bZ^{(b)})} \mid \bY^{(b)}}
  \leq \sum_{b=1}^B \lambda_b \log \esp[\btheta^{(h)}]{\frac{p_{\btheta^{(h+1)}}(\bY^{(b)}, \bZ^{(b)})}{p_{\btheta^{(h)}}(\bY^{(b)}, \bZ^{(b)})} \mid \bY^{(b)}}\,,
\end{align*} 
using Jensen's inequality, so
\begin{align*}
  0
  & \leq \sum_{b=1}^B \lambda_b \log \int_{\mathbb{R}^k} p_{\btheta^{(h)}}(\bZ^{(b)} \mid \bY^{(b)}) \frac{p_{\btheta^{(h+1)}}(\bY^{(b)}, \bZ^{(b)})}{p_{\btheta^{(h)}}(\bY^{(b)}, \bZ^{(b)})} \d \bZ^{(b)} \\ 
   & = \sum_{b=1}^B \lambda_b \log \left\lbrace \frac{1}{p_{\btheta^{(h)}}(\bY^{(b)})} \int_{\mathbb{R}^k} p_{\btheta^{(h+1)}}(\bY^{(b)}, \bZ^{(b)}) \d \bZ^{(b)} \right\rbrace 
  = \cl(\btheta^{(h+1)}) - \cl(\btheta^{(h)}). 
  \qquad \blacksquare
\end{align*} 

One may observe that this proof also holds for the decomposition \eqref{eqn:clZ}, replacing $\bZ^{(b)}$ with $\bZ$. As a consequence, replacing $\bZ^{(b)}$ with $\bZ$ in Algorithm \ref{algo:CL-EM} would yield the same property, although this alternative algorithm would not bring any computational advantage.

\subsubsection{Proof of Proposition \ref{prop:weight-norm}}
\label{proof:weight-norm}

\textbf{\emph{(i)}} For all $\bv\in\mathbb{R}^p$,
\begin{align*}
\frac{p_{\btheta^{(h)}}(\bY_i, \bv)^2}{\varphi(\bv; \bm, \bS)}
& =
\frac{\vert \bS\vert^{\nicefrac{1}{2}}}{(2\pi)^{\nicefrac{p}{2}}\left\vert \Sigma\right\vert}
\exp\left\lbrace -\frac{1}{2}\bv^\top\left(2\bSigma^{-1} - \bS^{-1}\right)\bv \right\rbrace 
\\
& \quad\quad\quad\times\exp\left[2\sum_{j=1}^p \left\lbrace Y_{ij}(o_{ij} + \bx_i^\top\bbeta^{(h)}_j + v_{j})  -\exp(o_{ij} + \bx_i^\top\bbeta^{(h)}_j + v_{j})  \right\rbrace\right].
\end{align*}
Thus, the latter is finite if the quadratic terms satisfy for all $\bv\in\mathbb{R}^p$,
\begin{equation*}
\bv^\top\left(2\bSigma^{-1} - \bS^{-1}\right)\bv \geq 0.
\end{equation*}
The result follows.

\noindent \textbf{\emph{(ii)}} For all $\bv\in\mathbb{R}^p$,
\begin{align} \label{eqn:weight-norm}
\frac{p_{\btheta}(\bY_i, \bv)}{\varphi(\bv\,;\,\bm, \bS)} 
& =
\frac{\vert \bS\vert^{\nicefrac{1}{2}}}{\left\vert \bSigma\right\vert^{\nicefrac{1}{2}}}
\exp\left\lbrace -\frac{1}{2}\bv^\top\left(\bSigma^{-1} - \bS^{-1}\right)\bv \right\rbrace 
\\.
&  \quad\quad\quad\times\exp\left[\sum_{j=1}^p \left\lbrace Y_{ij}(o_{ij} + \bx_i^\top\bbeta^{(h)}_j + v_{j})  -\exp(o_{ij} + \bx_i^\top\bbeta^{(h)}_j + v_{j})  \right\rbrace\right] \,.
\nonumber 
\end{align}
We thus have a finite upper bound if the quadratic terms satisfy for all $\bv\in\mathbb{R}^p$,
\begin{equation*}
\bv^\top\left(\bSigma^{-1} - \bS^{-1}\right)\bv \geq 0.
\end{equation*}

\subsubsection{Proof of Proposition \ref{prop:weight-mixt}}
\label{proof:weight-mixt}

For all $\bv\in\mathbb{R}^p$, we have
\begin{equation*}
q_i^{(h)}(\bv) \geq (1 - \alpha)\varphi(\bv; \bm_i^{(h)}, \bSigma^{(h)})\,,
\end{equation*}
and then,
\begin{equation*}
\frac{p_{\btheta^{(h)}}(\bY_i, \bv)^2}{q_i^{(h)}(\bv)} \leq \frac{1}{(1 - \alpha)}\frac{p_{\btheta^{(h)}}(\bY_i, \bv)^2}{\varphi(\bv; \bm_i^{(h)}, \bSigma^{(h)})}.
\end{equation*}
The result follows from statement \emph{(i)} of Proposition \ref{prop:weight-norm}, taking $\bS = \bSigma = \bSigma^{(h)}$.

\subsection{Update formulas (M-step)}
\label{app:updt-em}

\paragraph*{Objective function}
\begin{align*}
Q(\btheta \mid \btheta^{(h)}) = \esp[\btheta^{(h)}]{\log p_{\btheta}(\bY, \bZ) \mid \bY}
& = 
\sum_{i = 1}^{n}
\esp[\btheta^{(h)}]{
\log p_{\bSigma}(\bZ_i) + \log p_{\bBeta}(\bY_i \mid \bZ_i) \mid \bY_i
}.
\end{align*}
The above decomposition allows to perform optimization in $\bBeta$ and $\bSigma$ separately.

\paragraph*{Update for $\bBeta$} For all $(i, j)\in \lbrace 1, \ldots, n\rbrace \times \lbrace1, \ldots, p\rbrace$
\begin{align*}
\frac{\partial}{\partial \bbeta_{j}} \log p_{\bBeta}(\bY_i \mid \bZ_i)
& = \frac{\partial}{\partial \bbeta_{j}} 
\left\lbrace (\bo_i + \bBeta^\top\bx_i + \bZ_i)^\top\bY_i - \sum_{k = 1}^p \exp\left(o_{ik} + \bx_i^\top\bbeta_k + Z_{ik}\right) \right\rbrace
\\ 
& = \left\lbrace \bY_{ij} - \exp\left(o_{ij} + \bx_i^\top\bbeta_j + Z_{ij}\right)\right\rbrace \bx_i.
\end{align*}
The gradient in $\bbeta_j$ thus writes as
\begin{align*}
\nabla^{(h)}(\bbeta_j) 
& \triangleq \frac{\partial}{\partial \bbeta_{j}} Q(\btheta \mid \btheta^{(h)})
= \sum_{i = 1}^n \left\lbrace \bY_{ij} - \exp\left(o_{ij} + \bx_i^\top\bbeta_j\right) \esp[\btheta^{(h)}]{\exp\left(Z_{ij}\right) \mid \bY_{ij}}\right\rbrace \bx_i \,,
\end{align*}
and its self-normalised importance sampling estimate for a $N$-sample $(\bV_{i1}, \ldots, \bV_{iN})$ from $q_i^{(h)}$
\begin{align*}
\widehat{\nabla}^{(h)}(\bbeta_j)
& =  \sum_{i = 1}^n \left\lbrace \bY_{ij} - \exp\left(o_{ij} + \bx_i^\top\bbeta_j\right) \is[{q}_i^{(h)}]{\exp\left(Z_{ij}\right)} \right\rbrace \bx_i 
\nonumber
\\
& =  \sum_{i = 1}^n \left\lbrace \bY_{ij} - \exp\left(o_{ij} + \bx_i^\top\bbeta_j\right) \sum_{r= 1}^N w^{(h)}_{ir}\exp\left(V_{irj}\right) \right\rbrace \bx_i \,.
\label{eqn:grad-beta}
\end{align*}

\paragraph*{Update for $\bSigma$} For all $(i, j, k)\in\lbrace 1, \ldots, n\rbrace \times \lbrace 1, \ldots, p\rbrace \times \lbrace 1, \ldots, p\rbrace$,
\begin{equation*}
\frac{\partial}{\partial \sigma_{jk}} \log p_{\bSigma}(\bZ_i) = -\frac{1}{2}\frac{\partial}{\partial \sigma_{jk}} \left(\log\vert \bSigma\vert + \bZ_i^\top \bSigma \bZ_i \right).
\end{equation*}
Denote $\mathbf{E}_{jk}$ the matrix unit with value 1 at coefficient $(j,k)$ and $\delta_{jk}$ the Kronecker delta. Since $\bSigma$ is symmetric, 
\begin{equation*}
\frac{\partial}{\partial \sigma_{jk}} \log\vert \bSigma\vert = \tr\left(\bSigma^{-1} \frac{\partial}{\partial \sigma_{jk}} \bSigma \right) = \tr(\bSigma^{-1}\mathbf{E}_{jk}) + (1 - \delta_{jk})\tr(\bSigma^{-1}\mathbf{E}_{kj}) = (2 - \delta_{jk})\Sigma^{-1}_{jk},
\end{equation*}
with $\bSigma^{-1} = (\Sigma^{-1}_{jk})$.
Furthermore, again using that $\bSigma$ is symmetric
\begin{align*}
\frac{\partial}{\partial \sigma_{jk}} \bZ_i^\top\bSigma^{-1}\bZ_i 
& = \sum_{u = 1}^p  Z_{iu} \sum_{v = 1}^p Z_{iv} \frac{\partial}{\partial \sigma_{jk}} \Sigma^{-1}_{uv}
\\
& = - \frac{2 - \delta_{jk}}{2} \sum_{u = 1}^p  Z_{iu}\sum_{v = 1}^p Z_{iv} (\Sigma^{-1}_{uj} \Sigma^{-1}_{kv} + \Sigma^{-1}_{uk} \Sigma^{-1}_{jv})
 \\
& = (\delta_{jk} - 2) \sum_{u = 1}^p Z_{iu} \Sigma^{-1}_{uj}\sum_{v = 1}^p Z_{iv}\Sigma^{-1}_{kv}
\\
& = (\delta_{jk} - 2) \left[\bSigma^{-1} Z_i Z_i^\top\bSigma^{-1}\right]_{jk}.
\end{align*}
The update for $\bSigma$ is then solution of
\begin{equation*}
\frac{\delta_{jk} - 2}{2} \left[
n\bSigma^{-1} - \bSigma^{-1}\sum_{i = 1}^n \esp[\btheta^{(h)}]{\bZ_i\bZ_i^\top \mid \bY_i}\bSigma^{-1}
\right]_{jk} = 0.
\end{equation*}
This leads to, for the same $N$-sample $(\bV_{i1}, \ldots, \bV_{iN})$ from $q_i^{(h)}$
\begin{equation*}
\bSigma^{(h+1)} = \frac{1}{n} \sum_{i = 1}^n  \esp[\btheta^{(h)}]{\bZ_i\bZ_i^\top \mid \bY_i}
\quad\text{and}\quad
\widehat{\bSigma}^{(h+1)} = \frac{1}{n}\sum_{i=1}^n\sum_{r=1}^N w^{(h)}_{ir} \bV_{ir}\bV_{ir}^\top.
\end{equation*}

\subsection{Update formulas (CL-M-step)}
\label{app:updt-cem}

\paragraph*{Objective function}
\begin{align*}
Q_{\cl}(\btheta \mid \btheta^{(h)}) & =  \sum_{b=1}^C \lambda_b \esp[\btheta^{(b, h)}]{\log p_{\btheta}(\bY^{(b)}, \bZ^{(b)}) \mid \bY^{(b)}} 
\\
& = 
 \sum_{b=1}^C \lambda_b 
\sum_{i = 1}^{n}
\esp[\btheta^{(b, h)}]{
\log p_{\bSigma^{(b)}}(\bZ^{(b)}_i) + \log {p_{\bBeta^{(b)}}(\bY^{(b)}_i \mid \bZ^{(b)}_i)} \mid \bY^{(b)}_i
}\,.
\end{align*}

\paragraph*{Update for $\bBeta$} For all $(i, j, b)\in\lbrace 1, \ldots, n\rbrace \times \lbrace 1, \ldots, p\rbrace \times \lbrace 1, \ldots, C\rbrace$,
\begin{align*}
\frac{\partial}{\partial \bbeta_{j}} \log p_{\bBeta^{(b)}}(\bY^{(b)}_i \mid \bZ^{(b)}_i)
& = 
\begin{cases}
\left\lbrace \bY_{ij} - \exp\left(o_{ij} + \bx_i^\top\bbeta_j + Z^{(b)}_{ij}\right)\right\rbrace \bx_i, & \text{if } j \in \mathcal{C}_{b},
\\
\mathbf{0}, & \text{otherwise}.
\end{cases}
\end{align*}
If we denote $\mathcal{C}(j) = \lbrace b\in\lbrace 1, \ldots, C\rbrace \mid j \in \mathcal{C}_b\rbrace$ the set of indices of blocks that contain $j$, the gradient in $\bbeta_j$ writes as
\begin{align*}
\nabla^{(h)}(\bbeta_j) 
& \triangleq \frac{\partial}{\partial \bbeta_{j}} Q_{\cl}(\btheta \mid \btheta^{(h)})
\\
& = \sum_{i = 1}^n \left\lbrace \bY_{ij} - \exp\left(o_{ij} + \bx_i^\top\bbeta_j\right) \left( \sum_{b\in\mathcal{C}(j)}\lambda_b \esp[\btheta^{(h)}]{\exp\left(Z^{(b)}_{ij}\right) \mid \bY_{ij}} \right)\right\rbrace \bx_i \,,
\end{align*}
and its self-normalised importance sampling estimate for $N$-samples $(\bV^{(b)}_{i1}, \ldots, \bV^{(b)}_{iN})$ from $q_i^{(b, h)}$
\begin{align*}
\widehat{\nabla}^{(h)}(\bbeta_j)
& =  \sum_{i = 1}^n \left\lbrace \bY_{ij} - \exp\left(o_{ij} + \bx_i^\top\bbeta_j\right)  \left( \sum_{b\in\mathcal{C}(j)}\lambda_b \is[{q}_i^{(b, h)}]{\exp\left(Z^{(b)}_{ij}\right)}\right) \right\rbrace \bx_i 
\nonumber
\\
& =  \sum_{i = 1}^n \left[ \bY_{ij} - \exp\left(o_{ij} + \bx_i^\top\bbeta_j\right)  \left\lbrace \sum_{b\in\mathcal{C}(j)}\lambda_b \sum_{r= 1}^N w^{(b, h)}_{ir}\exp\left(V^{(b)}_{irj}\right) \right\rbrace \right] \bx_i \,.
\label{eqn:grad-beta-composite}
\end{align*}

\paragraph*{Update for $\bSigma$.} Unlike the full data framework, we do not have access to an estimate of $\bSigma$ and should resort as well to a gradient-based method. 
Denote $\boldsymbol{\Omega}^{(b)}$ the inverse matrix of $\bSigma^{(b)}$ and for $(j,k) \in \lbrace 1, \ldots, p\rbrace^2$, $\mathcal{C}(j,k) = \left\lbrace b \in \lbrace 1,\ldots, C\rbrace \mid j,k \in \mathcal{C}_b\right\rbrace$
\begin{align*}
\frac{\partial}{\partial \sigma_{jk}} Q_{\cl}(\btheta \mid \btheta^{(h)})
& = \sum_{b \in\mathcal{C}(j,k)} 
 \frac{(2-\delta_{jk})\lambda_b}{2} \left[
 \boldsymbol{\Omega}^{(b)} \sum_{i=1}^{n} \esp[]{\bZ_i^{(b)} {\bZ_i^{(b)}}^{\top}} \boldsymbol{\Omega}^{(b)} - n\boldsymbol{\Omega}^{(b)}
 \right]^{\blacksquare}_{jk} \,.
\end{align*}
where $[\bA]^{\blacksquare}$ stands for a $k\times k$ matrix $\bA$ that has been embedded into a $p\times p$ matrix $\mathbf{M} = (M_{jk})$ with $M_{jk}$ taking value 0 if the pair $(j,k)$ does not belong to the possible pair of species of block $\mathcal{C}_b$ and being equal to the coefficient of $\bA$ related to the pair $(j,k)$ otherwise. Each term of the gradient in $\bSigma$ can thus be estimated using self-normalised importance estimates
\begin{align*}
\widehat{\frac{\partial}{\partial \sigma_{jk}} Q_{\cl}(\btheta \mid \btheta^{(h)})}
& = \sum_{b \in\mathcal{C}(j,k)} 
 \frac{(2-\delta_{jk})\lambda_b}{2} \left[
 \boldsymbol{\Omega}^{(b)} \sum_{i=1}^{n} \is[{q}_i^{(b, h)}]{\bZ_i^{(b)}{\bZ_i^{(b)}}^{\top}}\boldsymbol{\Omega}^{(b)} - n\boldsymbol{\Omega}^{(b)}
 \right]^{\blacksquare}_{jk}
\\
& = \sum_{b \in\mathcal{C}(j,k)} 
 \frac{(2-\delta_{jk})\lambda_b}{2} \left[
 \boldsymbol{\Omega}^{(b)} \sum_{i=1}^{n} \left(\sum_{r = 1}^N w_{i}^{(b, r)} \bV_{ir}^{(b)} {\bV_{ir}^{(b)}}^{\top}\right) \boldsymbol{\Omega}^{(b)} - n\boldsymbol{\Omega}^{(b)}
 \right]^{\blacksquare}_{jk} \,.
\end{align*}

\subsection{About the observed Fisher information matrix}
\label{app:fim}

Regular models admits an alternative definition of the Fisher information matrix, namely
\begin{equation*}
\bI(\btheta) = -\esp[\btheta]{\nabla^2_{\btheta} \log p_{\btheta}(\bY)}. 
\end{equation*}
Similarly to the score version, one can derive a sample mean estimate where the computation of the Hessian is made possible, for some models, using the second Louis' formula
\begin{equation} 
\label{eqn:louis-2}
\nabla^2_{\btheta} \log p_{\btheta}(\bY)
 = \esp[\btheta]{\nabla^2_{\btheta} \log p_{\btheta}(\bY, \bZ) \mid \bY}
 + \var[\btheta]{\nabla_{\btheta} \log p_{\btheta}(\bY, \bZ) \mid \bY}. 
\end{equation}
Combining \eqref{eqn:louis-1} and \eqref{eqn:louis-2}, the resulting estimator writes as
\begin{equation*}
\widehat{\bI}_n(\btheta) - \frac{1}{n} \sum_{i = 1}^n 
\esp[\btheta]{\nabla^2_{\btheta} \log p_{\btheta}(\bY_i, \bZ_i) + \nabla_{\btheta} \log p_{\btheta}(\bY_i, \bZ_i)\{\nabla_{\btheta} \log p_{\btheta}(\bY_i, \bZ_i)\}^\top\mid \bY_i} \,.
\end{equation*}
Interestingly, this estimator forms a special instance of control variates estimator. Indeed, the term in the conditional expectation integrates to 0 with respect to the joint distribution $p_{\btheta}(\bY, \bZ)$ since it can be written as $\nabla^2_{\btheta} p_{\btheta}(\bY_i, \bZ)/p_{\btheta}(\bY_i, \bZ)$. The practical benefits of such an estimator may nonetheless be limited. To ensure a variance reduction compared to the observed Fisher information matrix, it requires introducing and tuning a control coefficient to weight the above sum, a task that might not be done without introducing biased or a computational burden.

\subsection{Additional simulation results} 

\subsubsection{Comparison with adaptative importance sampling in a Gaussian mixture family} \label{sec:comp-ais}

The adaptive version of Gaussian mixture proposal distributions can be achieved with the M-PMC algorithm of \cite{cappe2008adaptive}. Specifically for the PLN model, in order to update the proposal distribution at each iteration $h$ for each site $i$, and potentially each block $b$, we may consider their defensive option, incorporating a Gaussian distribution with a covariance matrix that fulfils condition \ref{item:is-finite-var} of Proposition \ref{prop:weight-norm} as defensive component, \textit{e.g.}, for $0 \leq \alpha < 1$ and $K\in\mathbb{N}^*$,
\begin{equation}
\label{eqn:ais-proposal}
\widetilde{q}^{(h)}_i = (1 - \alpha) \varphi\left(\cdot \,;\,\bm^{(h)}_i, \bSigma^{(h)}\right) + \alpha \sum_{k = 1}^K \eta_k \varphi\left(\cdot \,;\, \widetilde{\bm}^{(h)}_{ik}, \widetilde{\bS}^{(h)}_{ik}\right).
\end{equation}
If we set the proposal distribution to lie within the family of two-component Gaussian mixtures ($K = 1$), the method resumes to recursively updating the mean and the covariance matrix of the free component to minimize the Kullback-Leibler divergence between the conditional distribution of interest and the mixture distribution. 
The latter optimum differs from our solution, which solely aims at minimizing the Kullback-Leibler divergence between the conditional distribution of interest and the free component. 

Denote $\widetilde{w}^{(h)}_{ir}$, $1 \leq r \leq N$, the importance weights associated with a $N$-sample from the mixture proposal \eqref{eqn:ais-proposal} 
and $\mathcal{R}$ the relative efficiency in terms of the normalized perplexity of the importance weights, that is
\begin{equation*}
\mathcal{R} = 
\exp\left(-\sum_{r = 1}^N w^{(h)}_{ir}\log w^{(h)}_{ir} \right) \left/ \left(\exp\left(-\sum_{r = 1}^N \widetilde{w}^{(h)}_{ir}\log \widetilde{w}^{(h)}_{ir} \right)\right) \right.
\,.
\end{equation*}
Over a dataset of 1000 count matrices randomly draws from the PLN model \eqref{eqn:pln} with $n = 100$, $p = 2, 3, 5$ and $7$, and $d = 3$,
Table \ref{tabap:ais-n-perp} shows that the adaptive scheme converges towards a mixture for which the Shannon entropy of the associated normalized weight is equivalent to our proposal.
\begin{table}[!h]
\caption{Normalized perplexity relative efficiency $\mathcal{R}$ for $N = 5000$ draws. Quantiles of the relative efficiency observed over a dataset of a 1000 count matrices corresponding to 100 independent draws from the PLN model \eqref{eqn:pln} ($n = 100$ sites, $d = 3$ covariates) for 10 different random parameter configurations.}
{\phantomsection
\label{tabap:ais-n-perp}}
\begin{tabular*}{\columnwidth}{@{\extracolsep\fill}lrrrr@{\extracolsep\fill}}
\toprule
Number of species $p$ & 2  & 3 & 5 & 7 \\
\midrule
$5\%$-quantile &  0.997 & 0.994 & 0.990 & 0.981 \\
$95\%$-quantile &  1.123 & 1.119 & 1.067 & 1.036 \\
\botrule
\end{tabular*}
\end{table}

For the same dataset, whether the number of free mixture component in the adaptive scheme is $K =1$ or $K=2$,  we can also observe that the $5\%$ and $95\%$ quantiles of Monte Carlo estimates of the Kullback-Leibler divergence $\mathrm{KL}[\widetilde{q}^{(h)}_i \Vert q^{(h)}_i]$ is close to zero (Table \ref{tab:ais-kl}). In the context of the PLN model, the adaptive scheme applied to Gaussian mixture models proves to be of little use. It implies an additional computational burden and does not lead to significant improvement as the proposal $q^{(h)}_i$ is already close to the optimal adaptive version.
\begin{table*}[!h]
\caption{Kullback-Leibler divergence $\mathrm{KL}[\widetilde{q}^{(h)}_i \Vert q^{(h)}_i]$ for $N = 5000$ draws and mixture proposal $\widetilde{q}^{(h)}_i$ with $K = 1$ or $K = 2$ free components. Quantiles of the Monte Carlo estimates for a dataset of a 1000 count matrices corresponding to 100 independent draws from the PLN model \eqref{eqn:pln} ($n = 100$ sites, $d = 3$ covariates) for 10 different random parameter configurations.}\label{tab:ais-kl}
\tabcolsep=0pt
\begin{tabular*}{\textwidth}{@{\extracolsep{\fill}}lrrrrrrrr@{\extracolsep{\fill}}}
\toprule%
& \multicolumn{4}{@{}c@{}}{$K = 1$} & \multicolumn{4}{@{}c@{}}{$K = 2$} \\
\cmidrule{2-5}\cmidrule{6-9}%
Number of species $p$ & 2  & 3 & 5 & 7 & 2  & 3 & 5 & 7 \\
\midrule
$5\%$-quantile ($\times 10^{-3}$) &  4.28 & 1.77 & 2.90 & 5.87 & 6.27 & 4.77 & 7.98 & 13.7  \\
$95\%$-quantile ($\times 10^{-1}$) &  1.31 & 1.36 & 0.80 & 0.59 & 1.35 & 1.38 & 0.83 & 0.71 \\
\botrule
\end{tabular*}
\end{table*}

\subsubsection{Full likelihood inference} \label{sec:betaDistFL}

We present here additional results on the distribution of the full likelihood estimates of the regression coefficients $\beta_{\ell j}$.

\begin{figure}[h!]
  \begin{center}
    \includegraphics[width=.29\textwidth, trim=20 20 20 20, clip=]{\dirsim/PvalKS-FL-score-n100-d3-p10-parm1\simulParms} 
    \caption{Distribution of the $p$-values from the Kolmogorov–Smirnov test applied to the distribution of the standardized estimates $\tbeta_{\ell j}$ over the $M = 100$ simulations as a function of the number of species ($p = 5, \dots, 10$) for full-likelihood inference (FL). 
  Dotted red lines: $\alpha = 5\%$ significance threshold after Bonferroni correction (\textit{i.e.}, $\alpha/(dp)$).}
    \label{fig:KSpval-FL}
  \end{center}
\end{figure}

\begin{figure}[ht]
  \begin{center}
    \begin{tabular}{m{.075\textwidth}m{.22\textwidth}m{.22\textwidth}m{.22\textwidth}}
      & \multicolumn{1}{c}{$\beta_{1j}$} & \multicolumn{1}{c}{$\beta_{2j}$} 
      & \multicolumn{1}{c}{$\beta_{3j}$} 
      \\
      $p=5$ &
      \includegraphics[width=.2\textwidth, trim=20 20 20 20, clip=]{\dirsim/StatBeta1-2-n100-d3-p5-parm1\simulParms-FL} &
      \includegraphics[width=.2\textwidth, trim=20 20 20 20, clip=]{\dirsim/StatBeta2-2-n100-d3-p5-parm1\simulParms-FL} & 
      \includegraphics[width=.2\textwidth, trim=20 20 20 20, clip=]{\dirsim/StatBeta3-2-n100-d3-p5-parm1\simulParms-FL} 
      \\
      $p=6$ &
      \includegraphics[width=.2\textwidth, trim=20 20 20 20, clip=]{\dirsim/StatBeta1-2-n100-d3-p6-parm1\simulParms-FL} &
      \includegraphics[width=.2\textwidth, trim=20 20 20 20, clip=]{\dirsim/StatBeta2-2-n100-d3-p6-parm1\simulParms-FL} & 
      \includegraphics[width=.2\textwidth, trim=20 20 20 20, clip=]{\dirsim/StatBeta3-2-n100-d3-p6-parm1\simulParms-FL} 
      \\
      $p=7$ &
      \includegraphics[width=.2\textwidth, trim=20 20 20 20, clip=]{\dirsim/StatBeta1-2-n100-d3-p7-parm1\simulParms-FL} &
      \includegraphics[width=.2\textwidth, trim=20 20 20 20, clip=]{\dirsim/StatBeta2-2-n100-d3-p7-parm1\simulParms-FL} & 
      \includegraphics[width=.2\textwidth, trim=20 20 20 20, clip=]{\dirsim/StatBeta3-2-n100-d3-p7-parm1\simulParms-FL} 
      \\
      $p=8$ &
      \includegraphics[width=.2\textwidth, trim=20 20 20 20, clip=]{\dirsim/StatBeta1-2-n100-d3-p8-parm1\simulParms-FL} &
      \includegraphics[width=.2\textwidth, trim=20 20 20 20, clip=]{\dirsim/StatBeta2-2-n100-d3-p8-parm1\simulParms-FL} & 
      \includegraphics[width=.2\textwidth, trim=20 20 20 20, clip=]{\dirsim/StatBeta3-2-n100-d3-p8-parm1\simulParms-FL} 
      \\
      $p=9$ &
      \includegraphics[width=.2\textwidth, trim=20 20 20 20, clip=]{\dirsim/StatBeta1-2-n100-d3-p9-parm1\simulParms-FL} &
      \includegraphics[width=.2\textwidth, trim=20 20 20 20, clip=]{\dirsim/StatBeta2-2-n100-d3-p9-parm1\simulParms-FL} & 
      \includegraphics[width=.2\textwidth, trim=20 20 20 20, clip=]{\dirsim/StatBeta3-2-n100-d3-p9-parm1\simulParms-FL} 
      \\
      $p=10$ &
      \includegraphics[width=.2\textwidth, trim=20 20 20 20, clip=]{\dirsim/StatBeta1-2-n100-d3-p10-parm1\simulParms-FL} &
      \includegraphics[width=.2\textwidth, trim=20 20 20 20, clip=]{\dirsim/StatBeta2-2-n100-d3-p10-parm1\simulParms-FL} & 
      \includegraphics[width=.2\textwidth, trim=20 20 20 20, clip=]{\dirsim/StatBeta3-2-n100-d3-p10-parm1\simulParms-FL} 
    \end{tabular}
    \caption{qq-plots of the normalized regression coefficients $\tbeta_{\ell j}$, as defined in Equation~\eqref{eq:betaTilde}, for the second simulated species and each of the $d = 3$ covariates, using full likelihood inference (FL), with $p = 5$ to $10$ species.
    $x$-axis: standard normal quantiles, $y$-axis: quantiles of $\tbeta_{\ell j}$ (black dots [$\bullet$]), magenta dashed lines [\textcolor{magenta}{- -}]: $95\%$ bounds for the standard normal qq-plot.
    }
      \phantomsection
    \label{fig:qqplot-BetaTildeNbSpecies}
  \end{center}
\end{figure}

\FloatBarrier

\subsubsection{Composite likelihood inference} \label{sec:betaDistCL}

\paragraph*{Distribution of the estimates}
We provide here
additional results regarding the distribution of the composite likelihood estimates of the regression coefficients $\beta_{\ell j}$.
We used the quantile-quantile plot (qq-plot) to have more insight on the potential deviation from normality.
A good fit occurs when the empirical quantiles lie within the confidence interval delimited in magenta.
Figure \ref{fig:qqplot-BetaTildeP30} shows the good fit of the test statistics defined in Equation \eqref{eq:betaTilde} to the standard normal for all composite likelihood algorithms. 
These results strikingly contrast with those obtained for the specific case of full likelihood inference presented in Appendix \ref{sec:betaDistFL}. 
Figure \ref{fig:KSpval-FL} shows a systematic increase of the deviation from normality when the number of species grows from $p=5$ to $p=10$. 
The $p$-values drops even faster once $p\geq 7$.
The qq-plots from Figure \ref{fig:qqplot-BetaTildeNbSpecies} outlines that the FL algorithm, while deviating from the normal distribution, tends to over-estimate $\Var[\widehat{\beta}_{\ell j}]$ as $p$ increases. 
The discrepancy between composite likelihood and full likelihood approaches can be further observed on Figure \ref{fig:qqplotBetaTildeP10}. 
On that instance for $p=10$, the distribution of the $\widehat{\beta}_{\ell j}$ resulting from the different CL$k$ algorithms do fit a standard Gaussian, whereas those resulting from FL does not.


\paragraph*{Effect of the block size (Figure \ref{fig:RatioVarBeta})} ~

\begin{figure}[!h]
  \begin{center}
    \includegraphics[width=.4\textwidth, trim=10 10 25 25, clip=]{\dirsim/RatioVarBeta-score-n100-d3-p30-parm1\simulParms-refCL5} 
    \caption{Boxplots of the relative variance of the regression coefficient estimates $\widehat{\beta}_{\ell j}$ obtained with the composite likelihood method (CL$k$) with block sizes $k = 2, 3, 7$, as compared to the CL5 algorithm (namely, $\widehat{\Var}_{\text{CL}k}[\widehat{\beta}_{\ell j}] / \widehat{\Var}_{\text{CL5}}[\widehat{\beta}_{\ell j}]$), for $p = 30$ species. Each boxplot is built across the $d\times p = 90$ normalised coefficients $\tbeta_{\ell j}$.}
    \label{fig:RatioVarBeta}
  \end{center}
\end{figure}


\paragraph*{Computational time (Table \ref{tab:unitaryComputTime})}~~
\begin{table*}[!h]
    \caption{
    Computational time in seconds for one EM step in one block (averaged over the $M = 100$ simulations) for $p = 10, 30$, and 50 species for each of the composite likelihood algorithm CL$k$, with block sizes $k = 2, 3, 5$ and $7$.
    \label{tab:unitaryComputTime}
    }
\tabcolsep=1pt
\begin{tabular*}{\textwidth}{@{\extracolsep{\fill}}lrrrr@{\extracolsep{\fill}}}
\toprule%
Algorithm & CL$2$ & CL$3$ & CL$5$ & CL$7$ \\ 
\midrule
        $p = 10$ & 0.008 & 0.021 & 0.336 & 0.492 \\ 
        $p = 30$ & 0.031 & 0.220 & 0.041 & 0.435 \\ 
        $p = 50$ & -- & 0.046 & 0.119 & 0.371 \\
\botrule
\end{tabular*}
\end{table*}

\begin{figure}[!ht]
  \begin{center}
    \begin{tabular}{rm{.22\textwidth}m{.22\textwidth}m{.22\textwidth}}
      & \multicolumn{1}{c}{$\beta_{1j}$} & \multicolumn{1}{c}{$\beta_{2j}$} 
      & \multicolumn{1}{c}{$\beta_{3j}$} 
      \\
      CL2 &
      \includegraphics[width=.2\textwidth, trim=20 20 20 20, clip=]{\dirsim/StatBeta1-2-n100-d3-p30-parm1\simulParms-CL2} &
      \includegraphics[width=.2\textwidth, trim=20 20 20 20, clip=]{\dirsim/StatBeta2-2-n100-d3-p30-parm1\simulParms-CL2} & 
      \includegraphics[width=.2\textwidth, trim=20 20 20 20, clip=]{\dirsim/StatBeta3-2-n100-d3-p30-parm1\simulParms-CL2} 
      \\
      CL3 &
      \includegraphics[width=.2\textwidth, trim=20 20 20 20, clip=]{\dirsim/StatBeta1-2-n100-d3-p30-parm1\simulParms-CL3} &
      \includegraphics[width=.2\textwidth, trim=20 20 20 20, clip=]{\dirsim/StatBeta2-2-n100-d3-p30-parm1\simulParms-CL3} & 
      \includegraphics[width=.2\textwidth, trim=20 20 20 20, clip=]{\dirsim/StatBeta3-2-n100-d3-p30-parm1\simulParms-CL3} 
      \\
      CL5 &
      \includegraphics[width=.2\textwidth, trim=20 20 20 20, clip=]{\dirsim/StatBeta1-2-n100-d3-p30-parm1\simulParms-CL5} &
      \includegraphics[width=.2\textwidth, trim=20 20 20 20, clip=]{\dirsim/StatBeta2-2-n100-d3-p30-parm1\simulParms-CL5} & 
      \includegraphics[width=.2\textwidth, trim=20 20 20 20, clip=]{\dirsim/StatBeta3-2-n100-d3-p30-parm1\simulParms-CL5} 
      \\
      CL7 &
      \includegraphics[width=.2\textwidth, trim=20 20 20 20, clip=]{\dirsim/StatBeta1-2-n100-d3-p30-parm1\simulParms-CL7} &
      \includegraphics[width=.2\textwidth, trim=20 20 20 20, clip=]{\dirsim/StatBeta2-2-n100-d3-p30-parm1\simulParms-CL7} & 
      \includegraphics[width=.2\textwidth, trim=20 20 20 20, clip=]{\dirsim/StatBeta3-2-n100-d3-p30-parm1\simulParms-CL7}
      \\
      VEM &
      \includegraphics[width=.2\textwidth, trim=20 20 20 20, clip=]{\dirsim/StatBeta1-2-n100-d3-p30-parm1\simulParms-VEM} &
      \includegraphics[width=.2\textwidth, trim=20 20 20 20, clip=]{\dirsim/StatBeta2-2-n100-d3-p30-parm1\simulParms-VEM} & 
      \includegraphics[width=.2\textwidth, trim=20 20 20 20, clip=]{\dirsim/StatBeta3-2-n100-d3-p30-parm1\simulParms-VEM}
      \\
      JK &
      \includegraphics[width=.2\textwidth, trim=20 20 20 20, clip=]{\dirsim/StatBeta1-2-n100-d3-p30-parm1\simulParms-JK} &
      \includegraphics[width=.2\textwidth, trim=20 20 20 20, clip=]{\dirsim/StatBeta2-2-n100-d3-p30-parm1\simulParms-JK} & 
      \includegraphics[width=.2\textwidth, trim=20 20 20 20, clip=]{\dirsim/StatBeta3-2-n100-d3-p30-parm1\simulParms-JK}
    \end{tabular}
    \caption{
      qq-plots of the normalized regression coefficients $\tbeta_{\ell j}$, as defined in Equation~\eqref{eq:betaTilde}, for each of the $d = 3$ covariates and for the second simulated species (out of $p = 30$), obtained using the composite likelihood algorithms (CL$k$) with block of size $k = 2, 3, 5$, and $7$, and the variational EM (VEM) algorithm with jackknife variance estimates.
      Same legend as Figure \ref{fig:qqplot-BetaTildeNbSpecies}}
      \phantomsection
    \label{fig:qqplot-BetaTildeP30}
  \end{center}
\end{figure}

\begin{figure}[!h]
  \begin{center}
    \begin{tabular}{m{.05\textwidth}m{.22\textwidth}m{.22\textwidth}m{.22\textwidth}}
      & \multicolumn{1}{c}{$\beta_{1j}$} & \multicolumn{1}{c}{$\beta_{2j}$} 
      & \multicolumn{1}{c}{$\beta_{3j}$} 
      \\
      FL &
      \includegraphics[width=.2\textwidth, trim=20 20 20 20, clip=]{\dirsim/StatBeta1-2-n100-d3-p10-parm1\simulParms-FL} &
      \includegraphics[width=.2\textwidth, trim=20 20 20 20, clip=]{\dirsim/StatBeta2-2-n100-d3-p10-parm1\simulParms-FL} & 
      \includegraphics[width=.2\textwidth, trim=20 20 20 20, clip=]{\dirsim/StatBeta3-2-n100-d3-p10-parm1\simulParms-FL} 
      \\
      CL2 &
      \includegraphics[width=.2\textwidth, trim=20 20 20 20, clip=]{\dirsim/StatBeta1-2-n100-d3-p10-parm1\simulParms-CL2} &
      \includegraphics[width=.2\textwidth, trim=20 20 20 20, clip=]{\dirsim/StatBeta2-2-n100-d3-p10-parm1\simulParms-CL2} & 
      \includegraphics[width=.2\textwidth, trim=20 20 20 20, clip=]{\dirsim/StatBeta3-2-n100-d3-p10-parm1\simulParms-CL2} 
      \\
      CL3 &
      \includegraphics[width=.2\textwidth, trim=20 20 20 20, clip=]{\dirsim/StatBeta1-2-n100-d3-p10-parm1\simulParms-CL3} &
      \includegraphics[width=.2\textwidth, trim=20 20 20 20, clip=]{\dirsim/StatBeta2-2-n100-d3-p10-parm1\simulParms-CL3} & 
      \includegraphics[width=.2\textwidth, trim=20 20 20 20, clip=]{\dirsim/StatBeta3-2-n100-d3-p10-parm1\simulParms-CL3} 
      \\
      CL5 &
      \includegraphics[width=.2\textwidth, trim=20 20 20 20, clip=]{\dirsim/StatBeta1-2-n100-d3-p10-parm1\simulParms-CL5} &
      \includegraphics[width=.2\textwidth, trim=20 20 20 20, clip=]{\dirsim/StatBeta2-2-n100-d3-p10-parm1\simulParms-CL5} & 
      \includegraphics[width=.2\textwidth, trim=20 20 20 20, clip=]{\dirsim/StatBeta3-2-n100-d3-p10-parm1\simulParms-CL5} 
      \\
      CL7 &
      \includegraphics[width=.2\textwidth, trim=20 20 20 20, clip=]{\dirsim/StatBeta1-2-n100-d3-p10-parm1\simulParms-CL7} &
      \includegraphics[width=.2\textwidth, trim=20 20 20 20, clip=]{\dirsim/StatBeta2-2-n100-d3-p10-parm1\simulParms-CL7} & 
      \includegraphics[width=.2\textwidth, trim=20 20 20 20, clip=]{\dirsim/StatBeta3-2-n100-d3-p10-parm1\simulParms-CL7}
    \end{tabular}
    \caption{qq-plots of the normalized regression coefficients $\tbeta_{\ell j}$, as defined in Equation~\eqref{eq:betaTilde}, for each of the $d = 3$ covariates and for the second simulated species (out of $p = 30$), obtained using the full likelihood method (FL) or the composite likelihood method (CL$k$) with block sizes $k = 2, 3, 5$, and $7$. 
    Same legend as Figure \ref{fig:qqplot-BetaTildeNbSpecies}.}
    \label{fig:qqplotBetaTildeP10}
  \end{center}
\end{figure}

\end{document}